\documentclass[12pt,a4paper]{article}

\pdfoutput=1
\usepackage{amsmath}
\usepackage{amsfonts}
\usepackage{amssymb}
\usepackage{graphicx}
\usepackage{array}
\usepackage{cancel}
\usepackage{caption}
\usepackage{wrapfig}
\usepackage{mathrsfs}
\usepackage{cancel}
\usepackage{siunitx}
\usepackage{amsthm}
\usepackage{tikz-cd}
\usepackage{diagbox}
\usepackage{slashed}
\usepackage{relsize}
\usetikzlibrary{babel}

\usepackage[left=1.5cm,right=1.5cm,top=1.5cm,bottom=1.5cm,includefoot,footskip=1.5cm]{geometry}
\usepackage[OT1, T2A]{fontenc}
\usepackage[utf8]{inputenc}
\usepackage[english]{babel}
\usepackage[toc]{appendix}
\usepackage{csquotes}

\usepackage[sorting=none, citestyle=numeric-comp,
backend=biber
]{biblatex}
\usepackage{biblatex2bibitem}

\usepackage[colorlinks=true,
linkcolor=blue,citecolor=green,breaklinks=true]{hyperref}

\let\{=\lbrace
\let\}=\rbrace
\let\[=\lbracket
\let\]=\rbracket
\let\dg=\dagger

\let\lc = \varepsilon
\let\ep = \epsilon
\let\w = \omega

\newcommand\ili{\int\limits}
\newcommand {\lb} {\left(}
\newcommand {\rb} {\right)}

\newcommand\p{\partial}

\newcommand{\ket}[1]{\left| #1 \right\rangle}
\newcommand{\bra}[1]{\left\langle #1 \right|}
\newcommand{\braket}[2]{\langle #1|#2 \rangle}
\newcommand{\bl}{\left\langle}
\newcommand{\br}{\right\rangle}

\newcommand{\mt}[1]{\mathcal{#1}}
\newcommand{\mtf}[1]{\mathfrak{#1}}

\DeclareMathOperator{\Tr}{Tr}

\DeclareMathOperator{\Li}{Li}

\numberwithin{equation}{section}
\usepackage{authblk}

\begin{document}
\author{Ilia V. Kochergin}
\affil{Department of Physics, Princeton University, Princeton, NJ 08544, USA}
\title{$1/N$ Corrections in $\text{QCD}_2$: Small Mass Limit and Threshold States}
\maketitle

\begin{abstract}
    In this paper we investigate $1/N$ corrections to mesonic spectrum in $1+1$-dimensional Quantum Chromodynamics ($\text{QCD}_2$) with fundamental quarks using effective Hamiltonian method. We express the corrections in terms of 't Hooft equation solutions. First, we consider 2-flavor model with a heavy and a light quark. We show that, in contrast to some claims in earlier literature, the $1/N$ correction to the mass of the heavy-light meson remains finite when the light quark mass is taken to zero. Nevertheless, the corrections become significantly larger in this limit; we attribute this to the presence of massless modes in the spectrum. We also study the corrections to the lightest meson mass in 1-flavor model and show that they are consistent with recent numerical data, but not with the prediction coming from bosonization. Then we study low energy effective theory for 2 flavors. We show that the 3-meson interaction vertex correctly reproduces Wess-Zumino-Witten (WZW) coupling when both quarks become massless. This coupling does not change even if one of the quarks is massive. We employ Discretized Light Cone Quatization (DLCQ) to check the continuum results and show that the improved version can be used for small quark mass. Finally, we study the states associated with $1\to 2$ meson thresholds. Using degenerate perturbation theory, we show that when the decay is allowed by parity, the infinite $N$ theory has near-threshold bound states that mix one- and two-meson parts. They are $1/3$ two-meson and $2/3$ one-meson and the corrections to their masses have unusual scaling $\sim 1/N^{2/3}$.
\end{abstract}
\newpage
\tableofcontents
\newpage

\section{Introduction}
Quantum chromodynamics (QCD) in 1+1 dimensions has been extensively used as a toy model for a real-world QCD for the past few decades. Namely, 't Hooft showed that the spectrum $\text{QCD}_2$ with gauge group $SU(N)$ and $N_f$ quarks in fundamental representation is solvable in large $N$ limit \cite{tHooft:1974pnl} when $N_f$ is fixed. This theory exhibits quark confinement: integrating out the non-dynamical gauge field gives a linear interaction potential between quarks and the spectrum is color-singlet. It consists of mesonic and baryonic states. In this sense the theory is similar to $\text{QCD}_4$. Since then the infinite $N$ model was extensively studied \cite{Brower:1978wm, Chabysheva:2012fe, Callan:1975ps, Burkardt:2000ez,Lebed:2000gm,Fateev:2009jf,Ziyatdinov:2010vg,Vegh:2023snc,Ambrosino:2023dik} and several discretization methods were developed for the finite $N$ case \cite{Pauli:1985ps,Hornbostel:1988fb,Hornbostel:1988ne,Anand:2020gnn,Anand:2021qnd}. Some results in the infrared (IR) limit were obtained using bosonization techniques \cite{Steinhardt:1980ry,Frishman:1987cx,   Witten:1983ar,Frishman:1992mr,Delmastro:2021otj}. Calculation of $1/N$ corrections in QCD$_2$ using an effective light-cone Hamiltonian for mesons was introduced in \cite{Barbon:1994au}, and we develop their approach further. Another well-studied two-dimensional model for QCD is $SU(N)$ gauge theory coupled to an adjoint majorana fermion, sometimes referred to as adjoint $\text{QCD}_2$ \cite{Dalley:1992yy,Bhanot:1993xp,Kutasov:1993gq,Trittmann:2000uj,Dempsey:2021xpf,Asrat:2022aov,Dempsey:2022uie,Trittmann:2023dar,Trittmann:2024jkf}. It has many interesting properties, such as supersymmetry at a special value of the quark mass \cite{Bhanot:1993xp,Kutasov:1993gq}, but even at large $N$ one has to use numerical methods to find the spectrum. Only for very heavy adjoint fermions some analytical treatment becomes possible \cite{Asrat:2022aov}. In this paper we will focus on $1/N$ corrections to the infinite $N$ spectrum, so we will work with fundamental $\text{QCD}_2$ where it is known analytically.

The idea to study QCD in the limit of large number of colors was proposed by 't Hooft \cite{tHooft:1973alw}. In this limit with $g^2 N$ being fixed ($g$ is the Yang-Mills coupling) nonplanar Feynman diagrams are suppressed in $1/N$ so the theory simplifies. It often leads to major simplifications, a prominent example is the already mentioned fundamental $\text{QCD}_2$. However, it is unclear whether the theory at small $N$ would be qualitatively similar. For instance, in the case of adjoint $\text{QCD}_2$ the $1/N$ corrections are small \cite{Dempsey:2022uie}, which was recently studied analytically in \cite{Trittmann:2024jkf}. On the other hand, there are several indications that it might not be the case for very light fundamental quarks with mass $m\ll g$\footnote{Note that in 2d the coupling $g$ has a dimension of mass} (strong coupling). If $1/N$ expansion breaks down, we would need a basis of states different from the large $N$ one for perturbation theory. This can lead to significant changes in the spectrum. Nevertheless, our results imply that it does not happen for small quark masses.

The first sign that there can be issues with small $m$ is quark condensate $\bl\bar{\Psi}\Psi\br$. Large $N$ single flavor theory predicts $\bl\bar{\Psi}\Psi\br \neq 0$ at $m \to 0$ \cite{Zhitnitsky:1995qa}. Nonzero value means that the axial symmetry is spontaneously broken which is impossible in 2d due to Coleman's theorem \cite{Coleman:1973ci}. A nontrivial condensate can only appear at infinite $N$ as then the corresponding Goldstone boson is noninteracting. Besides, according to \cite{Zhitnitsky:1995qa}, $1/N$ corrections to some vacuum correlators blow up when $m \ll g$. Furthermore, in \cite{Barbon:1994au} it was found that the $1/N $correction to the heavy-light meson mass appears to diverge as $m\rightarrow 0$. Our calculations do not support the presence of this divergence and we prove the finiteness of this correction analytically, but we do find that it grows significantly at small $m$. This turns out to be related to the presence of a massless mode at $m = 0$~--- the feature absent in adjoint $\text{QCD}_2$.

Another issue with the small $m$ limit comes from bosonization. When the quarks are very light, one can write the low energy effective Lagrangian in terms of a bosonic field taking values in $U(N_f)$ group \cite{Witten:1983ar,Frishman:1992mr,Steinhardt:1980ry,Frishman:1987cx}. In the units of effective coupling $\sqrt{\frac{g^2 N}{2\pi}}$ for $N_f=1$ it predicts that the mass of the lightest mesonic state squared is $\mu^2 = \frac{4\pi e^\gamma}{\sqrt{N}}m$ when $N\to \infty$, where $\gamma$ is the Euler constant. Here $m$ is also given in the effective coupling units. On the other hand, 't Hooft model gives $\mu^2 = \frac{2\pi}{\sqrt{3}}m$ in this limit. There is a clear discrepancy between these results: the bosonization answer is parametrically smaller. If it was true, we would observe corrections at small $m$ that are not suppressed in $1/N$. Our results and the numerical computations of \cite{Anand:2021qnd} however seem to support the 't Hooft model formula.

One more result of bosonization is the low energy theory in the massless quark case. The full theory then has massless modes which are described by conformal $U(N_f)$ Wess-Zumino-Witten (WZW) model at level $N$. When $N_f = 1$ it is simply a free boson, but for higher $N_f$ the theory becomes interacting. The coupling is fixed by the current algebra. It vanishes at large $N$ and gives the three-meson vertex of the theory. For instance if $N_f = 2$ and the quarks are $q$ and $Q$, WZW model gives $\pi B \bar{B}$ derivative coupling where $\pi = q \bar{q}$ and $B = Q\bar{q}$ are the mesonic states. We will reproduce this coupling directly from the microscopic theory using $1/N$ corrections and show that surprisingly it does not change when $Q$ becomes massive. The effective theory for $B$ mesons is then a massive theory interacting with a conformal field theory (CFT) with coupling vanishing in the IR limit. It resembles an unparticle theory discussed by Georgi \cite{Georgi:2007ek}.

While the original way to compute $1/N$ corrections is based on diagrammatics, in 2d there is a more powerful Hamiltonian approach \cite{Pauli:1985ps,Barbon:1994au,Trittmann:2000uj,Thorn:1978kf}. It is based on quantization in lightcone coordinates $x^{\pm} = \frac{x^0\pm x^1}{\sqrt{2}}$ where $x^+$ is used as time. The gauge field can be integrated out so the resulting Hamiltonian would be purely in terms of the spinor fields. The main advantage of lightcone coordinates is nonnegativity of lightcone momenta $p^+$. It means that if the momentum is discretized by e.g. compactification of $x^-$ on a circle with radius $R$, there is only finite number of states with a given momentum. This allows one to diagonalize the Hamiltonian is a finite dimensional matrix and recover the spectrum~--- such method is called Discretized Lightcone Quantization (DLCQ). Unfortunately, it convergens very slowly for small quark masses. A way to improve the convergence for the large $N$ theory was proposed in \cite{vandeSande:1996mc} and we will see that it works rather well for finite $N$.

In this work we diagonalize the lightcone Hamiltonian at large $N$ and zero baryon number. This way we rewrite it in terms of mesonic creation and annihilation operators \cite{Barbon:1994au,Ji:2018waw}. Then at finite $N$ the Hamiltonian acquires multi-meson interaction terms which are suppressed in $1/N$. Due to momentum nonnegativity, the lightcone vacuum does not change when the interaction is turned on. It means that we can use the infinite $N$ states as a basis for the standard time-independent perturbation theory \cite{Thorn:1979gu}. The leading correction to a meson mass ($\sim 1/N$) is second order and comes from the three-point vertex. We study the corrections to the ground state $B$ meson mass in $N_f = 2$ model in the limit of very heavy $Q$\footnote{We take $N_f>1$ as otherwise the only potentially divergent correction vanishes from symmetry considerations}. Then all dynamical quantities including $1/N$ corrections do not depend on the heavy quark mass \cite{Georgi:1990um,Eichten:1989zv,Bigi:1997fj,Shifman:1995dn}. As we mentioned above, we see that the leading corrections remain finite even when the light quark mass goes to 0. It happens precisely because ground state pion $\pi_0$ decouples from the theory when it becomes massless. Moreover, in this case they are dominated by a single contribution coming from $\ket{B_0\pi_0}$ two-particle intermediate state where both mesons are ground states. It rapidly drops with an increase of the light quark mass.

We also show that one might encounter divergences at higher orders of perturbation theory for corrections to the lightest meson mass when the quark mass goes to $0$. However, if one carefully uses degenerate perturbation theory to resum the corrections the divergences disappear. Instead, some of the higher-order $1/N$ corrections are amplified, e.g. $1/N^2$ turns into $1/N$. There are also interesting cancellations between them. Nevertheless, it does not distort the infinite $N$ spectrum and the corrections to squared mass have the same $\sim m$ scaling in quark mass $m$ as the infinite $N$ meson mass squared $\mu_0^2$ at $m\to 0$. Hence they should always be subleading compared to $\mu_0^2$ and cannot change it very significantly. Nevertheless, the amplification of $1/N$ corrections means that the standard topological expansion for QCD \cite{tHooft:1973alw} is violated.

Finally, it turns out that besides the light quark limit there are other situations when the corrections might become singular. It happens when some $1\to 2$ meson decay threshold is exactly saturated. When the decay is allowed by parity\footnote{As meson ground states are odd, the decay $B_0\to B_0\pi_0$ is forbidden}, one has to use degenerate perturbation theory. It leads to restructuring of the infinite $N$ spectrum. Namely, for the $L\to R_1R_2$ threshold the bound state would be a combination of $\ket{L}$ and $\ket{R_1R_2}$ with non-vanishing 2-meson part at $N\to\infty$. Interestingly enough, the share of the 2-meson component in this state at large $N$ is $1/3$~--- it does not depend on the 3-point vertex structure. The correction to this bound state mass is $\sim 1/N^{2/3}$, which is an unusual scaling for $1/N$ corrections.

The rest of this work is organized as follows. In Section \ref{LQuant} we define the theory and quantize it in lightcone coordinates. We derive the Hamiltonian in terms of infinite $N$ mesonic operators and discuss important properties of the leading 3-point interaction. Then in Section \ref{CorrMes} we explain how to go to the heavy quark limit to separate the light quark dynamics. After that we present numerical results regarding $1/N$ corrections. We also analyze how our findings match with the calculations of \cite{Anand:2021qnd} and with the bosonization results. Section \ref{DLCQ} is dedicated to the use of improved DLCQ of \cite{vandeSande:1996mc} applied to $1/N$ corrections as an alternative to continuum methods. In the next Section \ref{EFT} we connect the meson interaction Hamiltonian with the low energy effective action to find the 3-meson IR coupling. Then we compare it with the WZW result. Lastly, in Section \ref{ThrSt} we explore the threshold states coming from the 3-meson vertex. We also discuss how the naive divergences one might encounter at higher orders of perturbation theory are resolved. We finish with a discussion of the results and possible future developments in section \ref{Disc}. The Appendices provide various technical details regarding numerical computations and various asymptotics.

\section{Lightcone quantization}\label{LQuant}
In this section we will lay the groundwork to study $1/N$ corrections in fundamental $\text{QCD}_2$. Note that we will use a different normalization of coupling constant $g$ than 't Hooft's original paper \cite{tHooft:1973alw}: $g = \sqrt{2}g_{\text{tH}}$.
\subsection{Hamiltonian formulation}
We start with the action of $SU(N)$ $\text{QCD}_2$ with $N_f$ flavors of fundamental quarks:
\begin{equation}
    S = \ili d^2 x \left[ -\dfrac{1}{2g^2}\Tr F_{\mu\nu} F^{\mu\nu} + \sum_{j=1}^{N_f} \bar{\Psi}_j(i\slashed{D} - m_j)\Psi_j \right],\quad \Psi_j = \dfrac{1}{2^{1/4}} \begin{pmatrix}
        \psi_j\\
        \chi_j
    \end{pmatrix}.\label{QCDAct}
\end{equation}
Here the covariant derivative acts in the usual way:
\begin{equation}
    D_\mu = \p_\mu - i A^a_\mu T^a
\end{equation}
and the $SU(N)$ structure constants $T^a$ are normalized as follows:
\begin{equation}
    \Tr T^a T^b = \dfrac{1}{2}\delta^{ab},\quad T^a_{\alpha\beta} T^a_{\gamma\delta} = \dfrac{1}{2} \left[\delta_{\alpha\delta}\delta_{\beta\gamma} - \dfrac{1}{N} \delta_{\alpha\beta}\delta_{\gamma\delta}\right].
\end{equation}
We also choose the following $\gamma$-matrices:
\begin{equation}
    \gamma^0 = \sigma_2,\quad \gamma^1 = i\sigma_1.
\end{equation}

As we mentioned in the introduction, we will use lightcone Hamiltonian approach to study this theory. First, we introduce the lightcone coordinates $x^{\pm} = \frac{x^0 \pm x^1}{\sqrt{2}}$ with $x^+$ being the lightcone time and pick the lightcone gauge $A_- = 0$. Then the fields $A_+$ and $\chi$ become non-dynamical and can be integrated out:
\begin{equation}
    (i\p_-)^2 A_+^a = \dfrac{g}{2}\sum_{j=1}^{N_f}\psi_j^\dagger T^a \psi_j,\quad i\p_-\chi_j = -\dfrac{m_j}{\sqrt{2}} \psi_j.\label{Constr}
\end{equation}
The resulting lightcone Hamiltonian $P^-$ is as follows:
\begin{equation}
    P^- = \ili d x^-\,T^{+-} = \dfrac{1}{2}\ili dx^- \left[ \sum_{j=1}^{N_f} m_j^2 \psi_j^\dagger \dfrac{1}{i\p_-} \psi_j + g^2 \sum_{j,k=1}^{N_f} \psi_j^\dagger T^a \psi_j \dfrac{1}{(i\p_-)^2} \psi_k^\dagger T^a \psi_k  \right],\label{LCHam}
\end{equation}
here $T^{\mu\nu}$ is the stress-energy tensor. Note that the second term gives the linear interaction potential. Integration over the zero mode of $A_\mu$ gives an additional color-singlet constraint on physical states:
\begin{equation}
    \ili dx^- \psi^\dagger_j T^a \psi_j \ket{\text{phys}} = 0.
\end{equation}

Next, we introduce the standard creation/annihilation operators:
\begin{equation}
    \psi_{j,\alpha} = \ili_0^{\infty} \dfrac{dp^+}{\sqrt{2\pi}} \left[ b_{j,\alpha}(p^+) e^{-i p^+ x^-} + d_{j,\alpha}^\dagger (p^+) e^{i p^+ x^-} \right],
\end{equation}
where $\alpha$ is a color index. The nontrivial commutation relations are as follows:
\begin{equation}
    \{b_{j,\alpha}(p^+),b^\dagger_{k,\beta}(q^+)\} = \{d_{j,\alpha}(p^+),d^\dagger_{k,\beta}(q^+)\} = \delta_{jk} \delta_{\alpha\beta} \delta(p^+-q^+).
\end{equation}
It is important that the lightcone momentum cannot be negative as $P^+ = \frac{P^0 + P^1}{\sqrt{2}}$ is nonnegative for physical states. In particular, it means that interactions do not change the Fock vacuum: due to momentum conservation it is impossible to have terms in the Hamiltonian containing only $b^\dagger$ and $d^\dagger$. There are some issues with modes at exactly zero momentum \cite{Heinzl:2000ht, Brodsky:1997de,Anand:2020gnn}, but for the infinite volume case they give measure zero contribution. In finite volume they can be avoided altogether by imposing antiperiodic boundary conditions as we do in the section \ref{DLCQ}.

As the spectrum consists of color-singlet states, it is convenient to define bosonic singlet operators:
\begin{equation}
    M^\dagger_{pp',jj'}= \dfrac{1}{\sqrt{N}}\sum_{\alpha} b^\dagger_{\alpha}(p,j) d^\dagger_{\alpha}(p',j'),~ B_{pp',jj'} =\sum_{\alpha} b^\dagger_{\alpha}(p,j) b_{\alpha}(p',j'),~D_{pp',jj'} =\sum_{\alpha} d^\dagger_{\alpha}(p,j) d_{\alpha}(p',j').\label{CompOp}
\end{equation}
Their commutation relations are presented in the Appendix \ref{aux}. We can then rewrite the Hamiltonian in terms of them. It can be split into three parts: $P^- = T+V+V_{\text{sing}}$. Here $T$ is the kinetic energy:
\begin{equation}
\begin{aligned}
    T &= \sum_j \dfrac{m_j^2}{2} \ili_0^\infty \dfrac{d k}{k}(B_{kk,jj} + D_{kk,jj})+\dfrac{g^2N}{8\pi}\sum_j \ili_0^\infty dk \lb\ili_0^\infty \dfrac{dp}{(k-p)^2} \rb(B_{kk,jj}+D_{kk,jj}) \\
    &-\dfrac{\bar{g}^2N}{8\pi}\sum_j \ili_0^\infty \dfrac{dk}{k} (B_{kk,jj}+D_{kk,jj}),\quad \bar{g}^2 = g^2 \lb1-\dfrac{1}{N^2}\rb.\label{KinEn}
\end{aligned}
\end{equation}
and $V$ is the potential energy containing all the interactions:
\begin{equation}
\begin{aligned}
    V &= -\dfrac{g^2 N}{8\pi} \sum_{i,i',j,j'}\ili dkdk'dp dp' \left[\mt{K}_{MM} M^\dagger_{kk',ii'} M_{pp',jj'} + \mt{K}_{MB} M^\dagger_{kk',ii'} B_{pp',jj'} + \mt{K}_{MD}M^\dagger_{kk',ii'} D_{pp',jj'} + \right.\\
    &+ \left. \mt{K}_{BD}B_{kk',ii'} D_{pp',jj'}+\mt{K}_{BB}B_{kk',ii'} B_{pp',jj'} + \mt{K}_{DD}D_{kk',ii'} D_{pp',jj'} + \text{h.c.}  \right],\label{IntPot}
\end{aligned}
\end{equation}
where h.c. stands for Hermitian conjugate and the kernels $\mt{K}$ are listed in the Appendix \ref{aux}. There is also a singular term $V_{\text{sing}}$ which appears as this Hamiltonian is not normal ordered:
\begin{equation}
    V_{\text{sing}} = \dfrac{g^2 N}{8\pi}\sum_j\dfrac{N_f}{N}\left( \ili dk_1 dk_2 \dfrac{\delta(k_1-k_2)}{(k_1-k_2)^2} \right) \ili dk (B_{kk,jj}+D_{kk,jj}).
\end{equation}
All of the integrals have limits $(0,+\infty)$. We also omitted the <<$+$>> superscripts signifying that we use lightcone momenta for simplicity.

The second term in the kinetic energy (\ref{KinEn}) has a strongly divergent integral over $dp$: the integrand has a quadratic divergence. It is a consequence of $A_+$ having a zero mode, so inversion  (\ref{Constr}) leads to divergences. However we will see that the interaction term makes this divergence more mild and it can be removed just by subtracting the zero mode. Besides, more careful treatment of this zero mode leads to a principal value regularization even when there is a quadratic divergence \cite{Anand:2021qnd}. 

Finally, color singlet states can be generated either by $M^\dagger_{kk',ii'}$ or by baryonic operators of the form $\mt{B}^\dagger = \ep^{\alpha_1\dots\alpha_n} b^\dagger_{j_1,\alpha_1}(p^+_1)\dots b^\dagger_{j_n,\alpha_n}(p^+_n)$. However, the baryon number $b$ is conserved and if we restrict our consideration to the sector with $b=0$ then every state can be generated by mesonic operators $M^\dagger$.

\subsection{Infinite $N$ basis and $1/N$-expansion}\label{InfNBas}
In order to construct $1/N$ expansion we need to define a basis of singlet states. As we noted before, mesonic states can be constructed by $M^\dagger_{kk'',ii'}$ operators. Their commutation relations are $[M_{kk',ii'},M_{pp',jj'}^\dagger] = \delta_{ij}\delta_{i'j'}\delta(k-p)\delta(k'-p') + O(1/N)$, meaning that they generate an orthonormal basis only in the limit $N\to \infty$. To resolve this issue we can use the $1/N$ expansion of the operator algebra (\ref{MBDAlg}) itself \cite{Ji:2018waw}:
\begin{equation}
\begin{aligned}
    M_{kk',ii'} &= M^{0}_{kk',ii'} -\dfrac{1}{2N} \sum_{l,l'} \ili dq dq' M^{0\dagger}_{qq',ll'} M^0_{kq',il'}M^0_{qk',li'} + O\lb\frac{1}{N^2} \rb;\\
    B_{kk',ii'} &= \sum_{l}\ili dq M^{0\dagger}_{kq,il} M^0_{k'q,i'l} + O\lb\frac{1}{N^2} \rb;\quad D_{kk',ii'} = \sum_{l}\ili dq M^{0\dagger}_{qk,li} M^0_{qk',li'} + O\lb\frac{1}{N^2}\rb.\label{OpExp}
\end{aligned}
\end{equation}
Here operators $M^0$ serve as infinite $N$ limit of $M$:
\begin{equation}
    \relax[M^0_{kk',ii'},M^{0\dagger}_{pp',jj'}] = \delta_{ij}\delta_{i'j'}\delta(k-p)\delta(k'-p'),
\end{equation}
so they correspond to an orthonormal basis of states.

We can now use $M^0$ to rewrite the Hamiltonian:
\begin{equation}
\begin{aligned}
    &~P^- = \dfrac{1}{2}\sum_{i,i'}\ili dk dk' \lb\dfrac{m_i^2}{k} + \dfrac{m_{i'}^2}{k'}\rb M^{0\dagger}_{kk',ii'} M^0_{kk',ii'} \\
    &-\dfrac{\bar{g}^2N}{4\pi}\sum_{i,i'} \ili dk dk' dp dp'\dfrac{\delta(k+k'-p-p')}{(k-p)^2}\lb M^{0\dagger}_{kk',ii'} M^0_{pp',ii'} - M^{0\dagger}_{kk',ii'}M^0_{kk',ii'}\rb\\
    &+\dfrac{g^2 N}{4\pi \sqrt{N}}\sum_{i,i',l}\ili dkdk' dpdp' dq\\
    &~\lb \dfrac{M^{0\dagger}_{kq,il}M^{0\dagger}_{pp',li'} M^0_{kk',ii'}}{(p+q)^2}\delta(p+p'+q-k') -\dfrac{M^{0\dagger}_{pp',il}M^{0\dagger}_{qk',li'} M^0_{kk',ii'}}{(p-k)^2}\delta(p+p'+q-k) + \text{h.c.} \rb + O\lb\frac{1}{N}\rb.
\end{aligned}\label{LcHam}
\end{equation}
We kept the higher-order terms in the quadratic part as they simply rescale the coupling. We can also compute the $1/N$ term in the Hamiltonian with (\ref{OpExp}), it is presented in the Appendix \ref{aux}. From this point we will set $\frac{\bar{g}^2 N}{2\pi} = 1$. {Note that more usual choice is to set $\frac{g^2 N}{2\pi} = 1$ instead \cite{Barbon:1994au,Anand:2021qnd}. In that case the coefficients at the interaction terms in the Hamiltonian have a simple form $\sim 1/N^{a}$ with $a$ being either integer or half-integer. For our choice this gets modified to $\frac{1}{1-1/N^2} \frac{1}{N^a}$. On the other hand, its main advantage is that the restriction of the Hamiltonian to single-meson states is diagonalized at finite $N$. For the most part of this work we will be interested in the leading order $O(1/N)$ corrections to the spectrum that arise from the 3-point vertex ($a=1/2$). Then the prefactor $\frac{1}{1-1/N^2}$ can be omitted as including $1/N^2$ yields a subleading contribution.}

We should start with a diagonalization of the quadratic part. To do that, we define meson creation and annihilation operators
\begin{equation}
    \mt{M}_{p,n,ii'}^\dagger = \sqrt{p} \ili_0^1 \phi_{n,ii'}(x) M^{0^\dagger}_{xp,(1-x)p|ii'},\quad [\mt{M}_{p,n,ii'},\mt{M}^\dagger_{k,m,jj'}] = \delta_{nm}\delta_{ij}\delta_{i'j'}\delta(p-k),
\end{equation}
here we assume that parton distribution functions or wave functions $\phi_{jj'}^{(n)}(x)$ are orthonormal on $(0,1)$ with respect to the standard $dx$ measure. The inverse expression is as follows:
\begin{equation}
    M_{pp',ii'}^{0\dagger} = \dfrac{1}{\sqrt{p+p'}}\sum_n \mt{M}^\dagger_{p+p',n,ii'} \phi^{(n)}_{ii'}\lb \dfrac{p}{p+p'}\rb.\label{MesOpInv}
\end{equation}
By requiring that the quadratic part should be diagonal we get the following eigenvalue equation on $\phi$:
\begin{equation}
\begin{aligned}
    \mu^{(n)2}_{ii'} \phi^{(n)}_{ii'}(x) &= \left[\dfrac{m_i^2}{x} + \dfrac{m_{i'}^2}{1-x}\right]  \phi^{(n)}_{ii'}(x)+ \mt{P}\ili_0^1 dy \dfrac{\phi^{(n)}_{ii'}(x)-\phi^{(n)}_{ii'}(y)}{(x-y)^2} \\
    &=\left[\dfrac{m_i^2-1}{x} + \dfrac{m_{i'}^2-1}{1-x}\right]\phi^{(n)}_{ii'}(x) - \mt{P} \ili_0^1 dy\dfrac{\phi^{(n)}_{ii'}(y)}{(x-y)^2}.\label{Th-Eq}
\end{aligned}
\end{equation}
Here $\mu^{(n)2}_{ii'}$ is the meson mass: an eigenvalue of the mass operator $\hat{\mu}^2 = 2 P^+ P^-$ in the infinite $N$ limit. Note that $P^+$ here acts as a number (the total lightcone momentum) due to the momentum conservation. As we claimed, the divergence of the integrand now is $\frac{1}{x-y}$ instead of $\frac{1}{(x-y)^2}$. The principal value regularization in this case can be obtained by assigning an infinitesimal mass to $A_+$ while inverting the constraints (\ref{Constr}). Alternatively one can discretize the momentum space and simply remove the zero mode as we do in Section \ref{DLCQ}. There is also an alternative form of the equation given by the second line of (\ref{Th-Eq}) which is sometimes more suitable for practical applications. The principal value in the latter case is defined as
\begin{equation}
    \mt{P} \dfrac{1}{x^2} = \dfrac{1}{2}\left[ \dfrac{1}{(x-i\ep)^2} + \dfrac{1}{(x+i\ep)^2} \right].
\end{equation}

This equation (\ref{Th-Eq}) is precisely the 't Hooft equation first derived in \cite{tHooft:1974pnl}. It has been extensively studied in the literature \cite{tHooft:1974pnl,Brower:1978wm, Chabysheva:2012fe, Callan:1975ps, Burkardt:2000ez,Lebed:2000gm,Fateev:2009jf,Ziyatdinov:2010vg,Vegh:2023snc,Ambrosino:2023dik}. Its spectrum is discrete and wave functions are orthogonal, however in a general situation an analytic solution is not known. When $m_i = m_i' \neq 0$ it is possible to get $\mu^{(n)}_{ii'}$ as an asymptotic series in $1/m_i$ \cite{Ziyatdinov:2010vg}. Nevertheless, one can obtain an asymptotic behavior of $\phi^{(n)}$ near the endpoints:
\begin{equation}
    \phi^{(n)}_{ii'}(x) \sim \begin{cases}
        x^{\beta_i},&x\to 0;\\
        (1-x)^{\beta_{i'}},&x\to 1,
    \end{cases}\label{PhiAsymp}
\end{equation}
where $\beta_i$ are defined as follows:
\begin{equation}
    \pi\beta_i \cot \pi\beta_i + m_i^2= 1.\label{BetEq}
\end{equation}
It means that for nonzero masses we have Dirichlet boundary conditions, but otherwise $\phi$ can be nonzero at the endpoints. Later we will see that one of the masses being zero leads to a Neumann boundary condition (zero derivative) at the corresponding endpoint. Besides, when $m_i=m_{i'} = 0$ there is an exact solution $\phi^{(0)}_{ii'} = 1$ with $\mu^{(0)}_{ii'} = 0$. In this situation the theory spectrum has a massless particle even at finite $N$.

We can now use (\ref{MesOpInv}) to rewrite $P^-$ in terms of the mesonic operators $\mt{M}$. Up to $O(1/N)$ we get:
\begin{equation}
\begin{aligned}
    &~P^- = \sum_{i,i'|n}\ili dp \dfrac{\mu^{(n)2}_{ii'}}{2p} \mt{M}^\dagger_{p,n,ii'} \mt{M}_{p,n,ii'} \\
    &+\dfrac{1}{2\sqrt{N}}\sum_{R_1,R_2,L} \ili \dfrac{dp_1 dp_2}{\sqrt{p_1p_2(p_1+p_2)}}\delta_{r_1l}\delta_{r_1'r_2}\delta_{r_2'l'}\mt{T}\lb L|R_1R_2;\dfrac{p_1}{p_1+p_2}\rb \left[\mt{M}^\dagger_{p_1,R_1} \mt{M}^\dagger_{p_2,R_2} \mt{M}_{p_1+p_2,L} + \text{h.c.}\right].
\end{aligned}\label{PEff}
\end{equation}
Here we used the notation of \cite{Barbon:1994au}: $R_1 = (r_1,r_1'|n_{R_1})$, $R_2=(r_2,r_2'|n_{R_2})$ and $L = (l,l'|n_L)$ define the meson states, the vertical bar signifies that flavor and state indices have different regions of summation. The vertex function $\mt{T}$ is as follows:
\begin{equation}
    \mt{T}(L|R_1R_2;z) = \dfrac{1-z}{z}\ili_0^1 dx dy \dfrac{\phi_{R_1}(1-x) \phi_{R_2}(y) \{ \phi_L[(1-x)z]  - \phi_L[ z+y(1-z)] \}}{\lb x + y \dfrac{1-z}{z} \rb^2}.\label{TVertGen}
\end{equation}
It allows us to compute the leading corrections to meson masses using the standard perturbation theory for the mass operator $\hat{\mu}^2$. In the case of single particle states the leading correction $\sim1/N$ comes from the second order: the $O(1/N)$ 4-point vertex (\ref{4PVert}) is normal-ordered and annihilates one-particle states. Hence, we get:
\begin{equation}
    \delta \mu_L^2 = \dfrac{1}{N}\sum_{R_1,R_2}\dfrac{\delta_{r_1l}\delta_{r_2'l'}\delta_{r_1'r_2}}{2^{\delta_{ll'}\delta_{r_1r_1'}}} \ili_0^1 dz \dfrac{[\mt{T}(L|R_1R_2;z)+\delta_{ll'}\delta_{r_1r_1'}\mt{T}(L|R_2R_1;1-z)]^2}{z(1-z)\left[\mu_L^2 - \frac{\mu_{R_1}^2}{z} - \frac{\mu_{R_2}^2}{1-z}\right]+i\ep} \equiv \dfrac{1}{N} \sum_{R_1,R_2} \Sigma_{L|R_1R_2}.\label{GenCorr}
\end{equation}
The $i\ep$ term here allows us to consider metastable states as well: a nonzero imaginary part of $\delta \mu_L ^2$ then would give the width of decay into two-particle continuum.

\subsection{Properties of the three-point vertex}\label{VertProp}
Before we proceed with actual computations, it is important to understand the key properties of the three-point vertex and the vertex function $\mt{T}$. Let us begin with the case of equal flavors $l=l'=r_1=r_1'=r_2=r_2'$ when the integrand of the correction (\ref{GenCorr}) gets an extra term. If $i=i'$ in the 't Hooft equation (\ref{Th-Eq}) then it becomes symmetric under $x\to 1-x$. It means that the wave functions are either symmetric or antisymmetric:
\begin{equation}
    \phi^{(n)}_{ii}(x) = (-1)^n \phi^{(n)}_{ii}(1-x).\label{SymPhi}
\end{equation}
It follows that
\begin{equation}
    \mt{T}(L|R_1R_2;z) = (-1)^{n_L+n_{R_1}+n_{R_2}+1} \mt{T}(L|R_2R_1;1-z),\label{SymTEq}
\end{equation}
the proof amounts to swapping the variables $x\leftrightarrow y$ in (\ref{TVertGen}) and using (\ref{SymPhi}). It means that for coincident flavors $\Sigma_{L|R_1R_2} = 0$ if $n_L+n_{R_1}+n_{R_2}$ is even.

Next, we move to a generic situation. We can find the asymptotic behavior of $\mt{T}$ using (\ref{PhiAsymp}):
\begin{equation}
    \mt{T}(L|R_1R_2;z)\sim \begin{cases}
        z^{\beta_{l}+\beta_{r_2}},&z\to0;\\
        (1-z)^{\beta_{l'}+\beta_{r_1'}}&z\to 1.
    \end{cases}
\end{equation}
It is interesting to see what happens if on of the quarks of $L$ becomes massless, say $m_{l'} = 0$ leading to $\beta_{l'} = 0$. A naive $z\to 1$ limit would be
\begin{equation}
    \mt{T}(L|R_1R_2;z) = (1-z)\ili_0^1 dx dy \dfrac{\phi_L(1-x)-\phi_L(1)}{x^2}\phi_{R_1}(1-x) \phi_{R_2}(y),\quad z\to 1.\label{TAsymp}
\end{equation}
From the limit $x\to 1$ of 't Hooft equation it follows that
\begin{equation}
    \ili_0^1 dx \dfrac{\phi_L(1-x) - \phi_L(1)}{x^2} = \left[m_l^2 - \mu_{ll'}^{(n_L)2} \right]\phi_L(1),\label{1Limit}
\end{equation}
as a principal value integral becomes a usual integral if the singularity is at one of the endpoints. From the convergence of this integral we get that $\phi_L'(1) = 0$~--- the Neumann boundary condition we mentioned before. Therefore the integral in (\ref{TAsymp}) is convergent as well. Hence,
\begin{equation}
   \mt{T}(L|R_1R_2;z)\sim 1-z,\quad z\to 1,~m_{l'}=0.\label{TAsympMassl}
\end{equation}
In the same way one can show that
\begin{equation}
    \mt{T}(L|R_1R_2;z)\sim z,\quad z\to 0,~m_{l}=0.
\end{equation}
This property will be important for corrections in the chiral limit when some of the quark masses become zero.

While (\ref{SymTEq}) generally does not hold for distinct flavors, we can use parity invariance of QCD to get another constraint. It is not very straightforward as the lightcone quantization explicitly breaks this symmetry which exchanges $P^-$ and $P^+$. The parity $P^{(n)}_{ii'}$ of single particle eigenstates should still be well defined. Indeed, one can find it by computing matrix elements with scalar and pseudoscalar densities \cite{Callan:1975ps, Burkardt:2000ez}. The expression is as follows:
\begin{equation}
    m_{i'} \ili_0^1 dx \dfrac{\phi^{(n)}_{ii'}(x)}{1-x} = -P^{(n)}_{ii'} m_i \ili_0^1 dx\dfrac{\phi^{(n)}_{ii'}(x)}{x}.
\end{equation}
It is not well-defined when one of the masses is zero so in that case we should take a limit. As the ground state wave function is sign-definite we obtain
\begin{equation}
    P^{(n)}_{ii'} = (-1)^{n+1}.
\end{equation}

Now we can attempt to apply parity invariance to the vertex function $\mt{T}(L|R_1R_2;z)$. In our formalism the mesons are on-shell and the lightcone momentum is conserved in a vertex while the energy is not. As parity transforms them into each other we need an energy conservation as well. It leads to a constraint on $z$:
\begin{equation}
    \mu_L^2 = \dfrac{\mu_{R_1}^2}{z} + \dfrac{\mu_{R_2}^2}{1-z},\label{OnShEq}
\end{equation}
it also gives the zeroes of the denominator in the integrand of (\ref{GenCorr}). Let us denote the roots by $z_{\pm}$ with $z_+ \ge z_-$, parity acts as $z\to \dfrac{\mu_{R_1}^2}{\mu_L^2 z}$ leading to $z_+ \leftrightarrow z_-$. Then $\mt{T}(L|R_1R_2;z)$ defines a scattering amplitude and the parity invariance implies the following relation:
\begin{equation}
    \mt{T}\lb L|R_1R_2;z_+\rb = (-1)^{n_L+n_{R_1}+n_{R_2}+1}\mt{T}\lb L|R_1R_2;z_-\rb.\label{SymTPar}
\end{equation}
One can easily see that it is consistent with (\ref{SymTEq}) in the equal flavor case when $R_1=R_2$ and consequently $z_- = 1-z_+$. It is unclear how to prove it directly from 't Hooft equation but we present some numerical evidence in the next section.

Finally, let us consider a situation when one of the mesons $L,R_1,R_2$ is massless (e.g. the corresponding quark masses are zero), we will denote it by $\pi_0$. First if $L=\pi_0$ then $\phi_L(x) = 1$ as we discussed before and from (\ref{TVertGen}) we get $\mt{T}(\pi_0|R_1R_2;z) \equiv 0$. In fact it is simply a consequence of $\pi_0$ being an eigenstate of the full Hamiltonian: the spectrum remains massless at finite $N$\footnote{Actually the finite $N$ massless state is defined as $\ili_0^1 dz\,M^\dagger_{zp,(1-z)p} \ket{0}$, but the series expansion (\ref{OpExp}) implies that one can replace $M^\dagger$ with $M^{0\dagger}$.} \cite{Anand:2021qnd,Delmastro:2021otj,Katz:2014uoa}. Next, let $R_2 = \pi_0$, then $z_+ = 1$, $z_- = \frac{\mu_{R_1}^2}{\mu_L^2}$. As we showed, $\mt{T}$ vanishes at endpoints so the parity relation (\ref{SymTPar}) yields $\mt{T}(L|R_1\pi_0;z_-) = 0$. Similarly, $\mt{T}(L|\pi_0R_2;z_+) = 0$~--- massless particles decouple on-shell. A derivation of this fact based on chiral symmetry can be found in \cite{Umeeda:2021llf}.

\section{Corrections to meson masses}\label{CorrMes}
Now we can concentrate on studying the corrections in more detail. We will focus on possible singularities of $\Sigma_{L|R_1R_2}$. The $i\ep$ prescription makes the integral in (\ref{GenCorr}) regular even if the denominator of the integrand has roots $z_{\pm}$ on $(0,1)$. It can only become singular when $z_+ = z_-$ leading to $\mu_L = \mu_{R_1}+\mu_{R_2}$~--- when $L$ is exactly at the threshold of decay into $R_1R_2$. For instance it is always the case when a massless flavor is present and, say, $L=R_1$, $R_2 = \pi_0$. We discuss other situation in the Section \ref{ThrSt}. As $n_L + n_{R_1}+n_{R_2} = 2 n_L$ is even, from our discussion in the subsection \ref{VertProp} the integrand trivially vanishes if all flavors coincide. For distinct flavors, we get
\begin{equation}
    \Sigma_{L|L\pi_0} = -\delta_{r_2'l'}\delta_{l'r_2}\ili_0^1 dz \dfrac{\mt{T}^2(L|L\pi_0;z)}{\mu_L^2(1-z)^2}.\label{SigSing}
\end{equation}
We see that $r_2'=r_2$. In \cite{Barbon:1994au} this correction was found to be divergent in the limit $m_{r_2}\to 0$. However, $m_{l'} = m_{r_2'} = 0$ so from (\ref{TAsympMassl}) $\mt{T}(L|L\pi_0;z) \sim 1-z$, $z\to 1$. In the Appendix \ref{DerCalc} we show that
\begin{equation}
\mt{T}'(L|L\pi_0;1) = \mu_L^2.\label{VertDer}
\end{equation}
It means that the above integral is actually convergent and the result of \cite{Barbon:1994au} is likely a numerical error. It is rather interesting that this derivative has such a simple form, in Section \ref{EFT} we discuss the relation of this result to effective field theory couplings. One can also consider the case $L=R_2$, $R_1 = \pi_0$, but it will be completely analogous up to flipping $z\to 1-z$.

Another consequence is that in the limit $m_l \to 0$ we get $\Sigma_{L|L\pi_0} \sim \mu_L^2$. This fact is important: it shows that we can take the limit $m_{l,l'} \to 0$ while keeping $N$ finite. According to the results of \cite{Zhitnitsky:1995qa}, corrections to some vacuum correlators computed in the large $N$ limit blow up when $m_l \ll g \sim \frac{1}{\sqrt{N}}$ (strong coupling regime). For instance, infinite $N$ results lead to a nonzero value of the chiral condensate $\bl \bar{\Psi} \Psi \br$ at $m_l = 0$, while at finite $N$ it is impossible due to Coleman's theorem \cite{Coleman:1973ci}. We see that nevertheless the meson spectrum does not qualitatively change. However we will show below that in the chiral limit the corrections become noticeably larger.

In order to investigate $\Sigma_{L|R_1R_2}$ more carefully we consider a model with two quark flavors $q$ and $Q$ just as in \cite{Barbon:1994au}. We denote $m = m_q$, $M = m_Q$ and use the following notation for the mesons: $\pi = q\bar{q}$, $B = Q\bar{q}$, $\bar{B} = q\bar{Q}$, $\eta = Q\bar{Q}$, then the only choice for $L$ is $B_n$. Bars are used to distinguish quarks from antiquarks. In order to focus on the light quark effects which cause the singularity at $z = 1$ we will take the limit $M\to \infty$ while keeping $m$ finite~--- then the heavy quark dynamics becomes universal. We will finish this section with discussing the $N_f = 1$ case studied in \cite{Anand:2021qnd} and the bosonization result for the lightest meson mass.

\subsection{Heavy-light limit}\label{HLlimit}
The heavy quark limit $M\to \infty$ has a wide range of applications in QCD \cite{Georgi:1990um,Eichten:1989zv,Bigi:1997fj,Shifman:1995dn}. Most notably, after removing the kinematic part coming from the mass $M$ itself, all physical quantities become independent on $M$. They depend only on the light quark physics. In two dimensions we can see it from the 't Hooft equation for $B$ mesons. Namely, we take $m_i = M$, $m_{i'} = m$ in (\ref{Th-Eq}) defining $t = M(1-x)$ and $g^{(n)}(t) = \phi_{B_n}(x)/\sqrt{M}$, due to this additional rescaling $g^{(n)}(\w)$ are orthonormal with respect to $d\omega$ integration. In the heavy quark limit $t\in(0,\infty)$, then by keeping the leading terms in $M$ we get:
\begin{equation}
\begin{aligned}
    \xi_n g^{(n)}(t) &= \left[ t+ \dfrac{m^2}{t} \right] g^{(n)}(t) + \mt{P} \ili_0^{+\infty} ds \dfrac{g^{(n)}(t)-g^{(n)}(s)}{(t-s)^2}\\
    &=\left[ t+ \dfrac{m^2-1}{t} \right] g^{(n)}(t) - \mt{P} \ili_0^{+\infty} ds \dfrac{g^{(n)}(s)}{(t-s)^2},\quad \xi_n = \lim_{M\to\infty}\frac{\mu^{2}_{B_n} - M^2}{M}.
\end{aligned}\label{ThEqHl}
\end{equation}
As we see, $\xi_n$ depends only on the light quark mass and $\mu_B^{(n)} = M + \xi_n/2 + O(1/M)$~--- after subtracting the kinematic term $M$ all dependence on $M$ disappears. A systematic way to compute $1/M$ corrections to $\mu_{B_n}$ can be found in \cite{Burkardt:2000ez}, the leading term is as follows:
\begin{equation}
    \mu_{B_n}^{2} = M^2 + M \xi_n + \dfrac{\xi_n^{2}-2}{3} + O\lb \dfrac{1}{M} \rb.
\end{equation}

\begin{figure}[ht!]
    \centering
    \includegraphics[width = 0.7\textwidth]{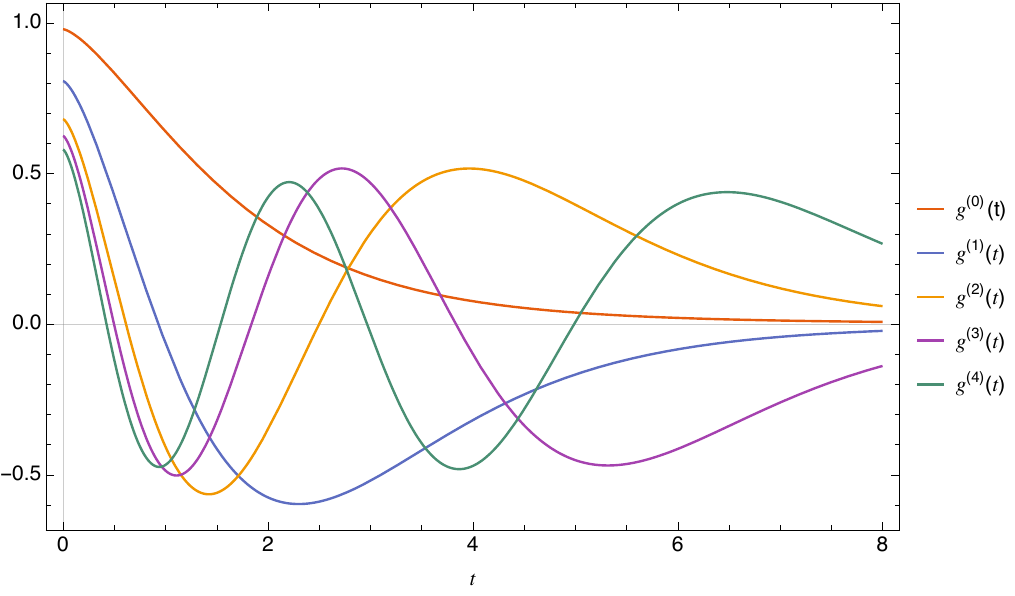}
    \caption{The first five wave functions for the $B$ meson states in the chiral limit}
    \label{InfFunc}
\end{figure}
A somewhat similar equation appeared in \cite{Donahue:2022jxu} for the case of scattering in the zigzag model, but the authors were unable to obtain an exact solution. We solve (\ref{ThEqHl}) numerically using an orthonormal basis approach explained in the Appendix \ref{HLThEq}. The first five eigenvalues $\xi_n$ for the chiral limit $m = 0$ are listed in Table \ref{EValTab}; Figure \ref{InfFunc} shows the corresponding wave functions $g^{(n)}(t)$. Similarly to the subsection \ref{VertProp} one can show that $g^{(n)'}(0) = 0$ when $m = 0$ which is noticeable on the Figure.
\begin{table}[h!]
\begin{center}
\begin{tabular}{|c||*{5}{c|}}
\hline
     n&0&1&2&3&4 \\
     \hline
    $\xi_n$&1.642286&4.303507&6.081913&7.476654&8.664182\\
    \hline
\end{tabular}
\caption{The first five eigenvalues $\xi_n$ in the chiral limt}\label{EValTab}
\end{center}
\end{table}

The vertex correction $\Sigma_{L|R_1R_2}$ from (\ref{GenCorr}) also admits a heavy quark limit when $L = B_n$, $R_1 = B_l$, $R_2 = \pi_k$. We denote $\omega = M(1-z)$, $\mu_k = \mu_{\pi_k}$ and $\phi^{(k)}(x) = \phi_{\pi_k}(x)$ (pion mass and pion wave function respectively), then 
\begin{equation}
    \Sigma_{B_n|B_l\pi_k} = -M \ili_0^{+\infty} d\w \dfrac{T_{n|lk}^2(\w)}{\w^2 + (\xi_l-\xi_n)\w + \mu_k^2 - i\ep}\equiv -M S_{n|lk},\label{SCorr}
\end{equation}
where $T_{n|lk}(\w)$ represents a rescaled $M\to \infty$ limit of $\mt{T}(B_n|B_l\pi_k;z)$:
\begin{equation}
    T_{n|lk}(\w) = \lim_{M\to\infty} \dfrac {\mt{T}(B_n|B_l\pi_k;z)}{M}=\omega \ili_0^{+\infty} dt \ili_0^1 dy \dfrac{g^{(l)}(t) \phi^{(k)}(y)\{g^{(n)}[t+\omega] - g^{(n)}[\omega(1-y)]\}}{(t+\w y)^2}\label{HLTVert}
\end{equation}
Note that the $M$ scaling is the same as in the heavy-light 't Hooft equation (\ref{ThEqHl}) so the $1/N$ corrections lead to an $M$-independent shift of the $B$-meson mass. We added a <<$-$>> sign to $S$ as we will mostly consider correction to the ground state which are negative. In order to compute it, besides $g^{(l)}$ and $g^{(n)}$ we need the pion wave functions $\phi^{(k)}$. The ways to find it numerically are given in the Appendices \ref{MasslThEq} (for the chiral limit $m=0$) and \ref{MassThEq} (for $m \neq 0$). The computation of $T$ is given in the Appendix \ref{HLVFunc}.

\begin{figure}[ht]
\begin{minipage}[h]{0.49\linewidth}
\center{\includegraphics[width=1\textwidth]{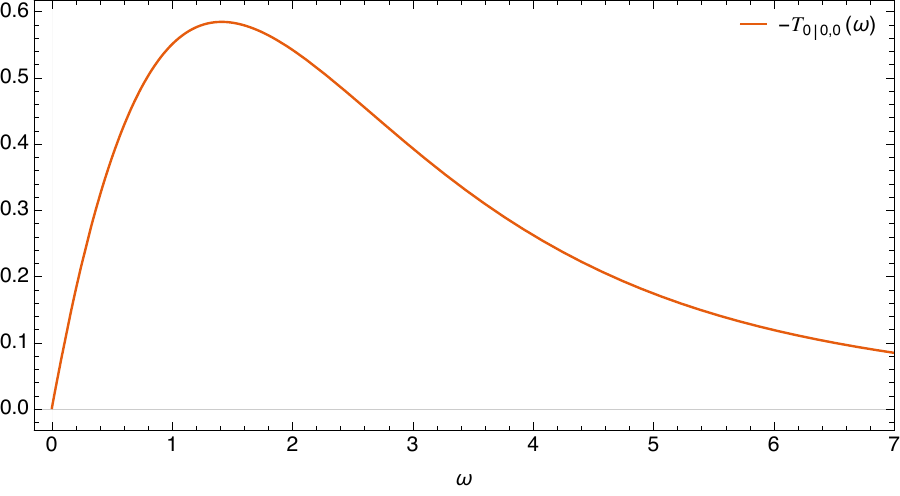} (a): $m=0$}
\end{minipage}
\begin{minipage}[h]{0.49\linewidth}
\center{\includegraphics[width=1\textwidth]{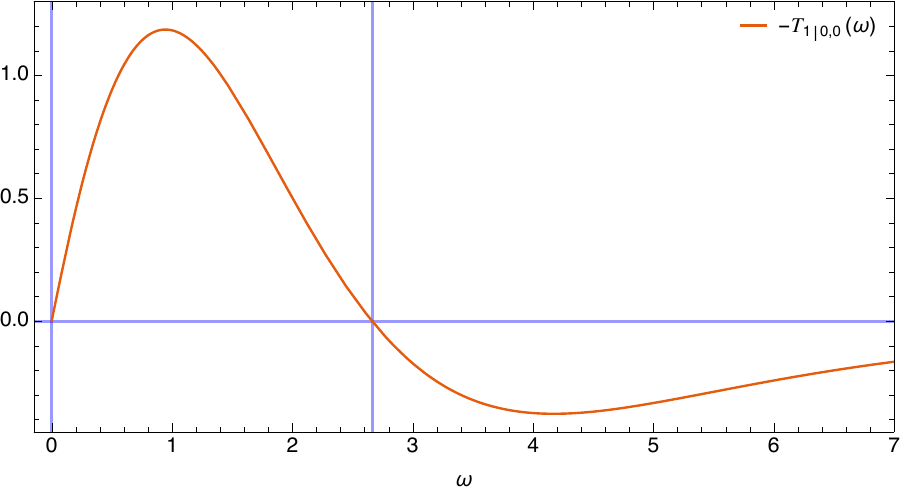} (b): $m = 0$}
\end{minipage}
\hfill

\begin{minipage}[h]{0.49\linewidth}
\center{\includegraphics[width=1\textwidth]{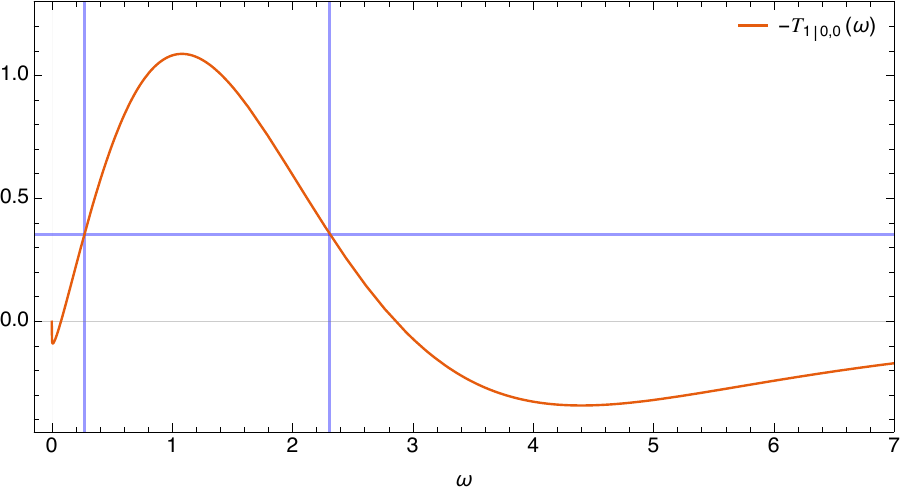} (c): $m = 0.15$}
\end{minipage}
\begin{minipage}[h]{0.49\linewidth}
\center{\includegraphics[width=1\textwidth]{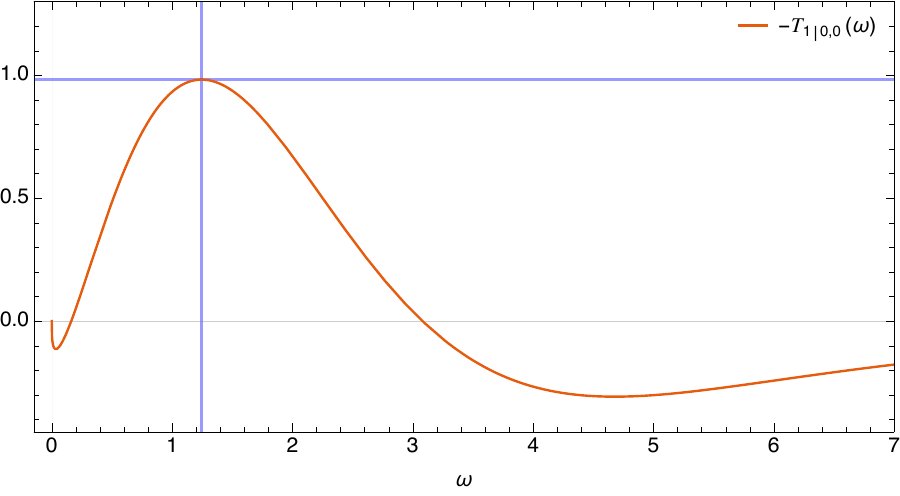} (d): $m=m_t$}
\end{minipage}
\caption{Vertex functions for different values of mass: (a)  $T_{0|0,0}(\w)$ for $m=0$; (b) $T_{1|0,0}(\w)$ for $m=0$; (c) $T_{1|0,0}(\w)$ for $m=0.15$; (d) $T_{1|0,0}(\w)$ for the threshold case $m=m_t$. The blue lines on (b), (c) and (d) show the on-shell roots $\w_{\pm}$ (vertical) and the corresponding values of $T_{1|0,0}$ (horizontal). The masses are given in terms of units with $\frac{\bar{g}^2 N}{2\pi} = 1$}\label{VertFunPlt}
\end{figure}
Figure \ref{VertFunPlt} shows the vertex functions $T_{0|0,0}(\w)$ for $m=0$ on (a) and $T_{1|0,0}(\w)$ on (b), (c), (d) for different values of $m$, we see that they drop at large $\omega$. We used the latter vertex to check the parity relation (\ref{SymTPar}), in the heavy-light case the on-shell points are as follows:
\begin{equation}
    \omega_{\pm} = \dfrac{\xi_n-\xi_l \pm \sqrt{(\xi_n-\xi_l)^2 - 4\mu_k^2}}{2}.\label{OnShHL}
\end{equation}
They are represented by grey vertical lines, and the corresponding values of $T_{1|00}$ by horizontal lines. (d) shows the threshold case $\xi_1-\xi_0 = 2\mu_0$ when $\w_{\pm}$ coincide, the corresponding $q$ mass is $m_t = 0.3248$. We see that the symmetry relation indeed holds: for instance $\xi_+$ is a zero of $T_{1|00}$ when $m = 0$. If $\w_{\pm}$ coincide parity implies that they should correspond to an extremum point, which we see on (d).

Finally, the denominator singularity in (\ref{SigSing}) corresponds to the case $l=n$, $k=0$ and $m=0$:
\begin{equation}
    \left. S_{l|l0}\right|_{m=0} = \ili_0^{+\infty} d\omega \dfrac{T^2_{l|l0}(\w)}{\w^2}.\label{SSing}
\end{equation}
Using the definition of $T$ as a limit (\ref{HLTVert}) and the derivative of the vertex function (\ref{VertDer}) we find that
\begin{equation}
    \left. T'_{n|n0}(0)\right|_{m=0} = -1.
\end{equation}
It is interesting that this result is rather simple and $n$-independent. The convergence of (\ref{SigSing}) trivially follows.

\subsection{Numerical results}\label{NumResSec}
In this subsection we discuss the result of numerical compuations of $1/N$ correction to the $B_0$ mass. We choose the ground $B$ meson state specifically so that the denominator of the integrand in (\ref{SCorr}) is always positive as the ground state cannot decay. In fact in the chiral limit $B_1$ and $B_2$ do not decay as well: from the Subsection \ref{VertProp} any decay amplitude containing a massless pion is zero; higher $B$ states can have decays involving $\pi_1$.

From (\ref{GenCorr}) and (\ref{SCorr}) it follows that the correction to $\mu_{B_0}$ is given by
\begin{equation}
    \delta \mu_{B_0}^2 = -\dfrac{M}{N} \sum_{l,k} S_{0|lk}.\label{HLcorr}
\end{equation}
In principle (\ref{GenCorr}) also yields a contribution from $\Sigma_{B_0|\eta_l B_k}$. However $\eta_l$ with the kinematic mass $2M$ is much heavier than $B_0$ so we can expect such terms to be negligible. We verify that in the Section \ref{DLCQ}. The values of $S_{0|lk}$ for $m = 0$ are presented in Table \ref{NumRes}. We see that $S_{0|0,0}$ is about two orders of magnitude larger than the subsequent terms: while the ground state contribution (\ref{SSing}) is indeed finite, it constitutes about 90\% of the correction. The contributions from higher excited states quickly drop with the increase of $k$ and $l$. The summation gives:
\begin{equation}
    \delta \mu_{B_0}^2 = -0.86 \dfrac{M}{N}.\label{DelMuBCont}
\end{equation}
For small $N$ this term is comparable to $\xi_0 M$~--- the dynamical large $N$ contribution to $\mu_{B_0}^2$. It is very different from the case of finite $N$ adjoint $\text{QCD}_2$ studied in \cite{Dempsey:2022uie,Trittmann:2024jkf}. That theory remains gapped even when the quark mass is zero and the coefficients of $1/N$-corrections (with the leading term $\sim N^{-2}$) are of order $10^{-2}$.
\begin{table}[ht!]
    \centering
    \begin{tabular}{|c|c|c|c|c|c|}
    \hline
         \diagbox{$l$}{$k$}&0&1&2&3&4  \\
         \hline
         0&0.818 & 0.0128 & 0.00174 & 0.000677 & 0.000258\\
         \hline
         1&0.00980 & 0.00422 & 0.000789 & 0.000306 & 0.000142\\
         \hline
         2&0.00615 & 0.00140 & 0.000403 & 0.000164 & 0.0000815\\
         \hline
         3& 0.00160 & 0.000994 & 0.000271 & 0.000117 & 0.0000604 \\
         \hline
         4&0.00159 & 0.000563 & 0.000195 & 0.0000857 & 0.0000460\\
         \hline
    \end{tabular}
    \caption{$S_{0|lk}$ in the chiral limit for the first five states}
    \label{NumRes}
\end{table}

Next, we investigate what happens with $S_{0|0,0}$ when we turn on $m$. It is instructive to understand the small $m$ asymptotic behavior first. As we show in the Appendix \ref{DerCalc},
\begin{equation}
    T_{0|0,0}(z) = -z + \dfrac{\pi}{\sqrt{3}}g^{(0)2}(0)\,m,\quad z\to 0,~|m \log z| \ll 1.
\end{equation}
From (\ref{SCorr}) it follows that
\begin{equation}
    S_{0|0,0}(m) = \mt{S}_0 - \sqrt{\dfrac{\pi^3 m}{2\sqrt{3}}} + \dfrac{\pi}{\sqrt{3}} g^{(0)^2}(0)\,m\log m + O(m),\quad m\to 0,\label{SAsymp}
\end{equation}
where we denoted $\mt{S}_0 = S_{0|0,0}(0)$. Here we used the well-known asymptotic form of $\mu_0^2$ at small $m$ \cite{Brower:1978wm,Callan:1975ps}:
\begin{equation}
    \mu_0^2 = \dfrac{2\pi}{\sqrt{3}}m,\quad m\to 0.\label{MuAsymp}
\end{equation}

This expression obviously cannot be used for sufficiently large $m$ as the correction coefficient $S_{0|0,0}$ is always positive. To get a more uniform result we can use the exact integral expression (\ref{SCorr}) and try to approximate $T_{0|0,0}$. Judging by the form of $T_{0|0,0}$ on Figure \ref{VertFunPlt} it would be natural to assume the contribution to the integral mostly comes from the region near $\w = 0$. We cannot directly use the asymptotic form $-\w$ as the integral would diverge. The simplest way to resolve it is to introduce a cutoff: $T_{0|0,0}^{\text{appr}}(\w) = -\w\,\theta( \mt{S}_0-\omega)$. Note that this expression does not take the $m$ dependence of $T_{0|0,0}$ into account, below we will see that it still gives rather good results. The approximate correction coefficient is as follows:
\begin{equation}
    \mt{S}(m) = \ili_0^{\mt{S}_0}d\w \dfrac{\w^2}{\w^2+\mu_0^2(m)} = \mt{S}_0 - \mu_0(m) \arctan\dfrac{\mt{S}_0}{\mu_0(m)}.\label{SAppr}
\end{equation}
This formula correctly reproduces first two terms in (\ref{SAsymp}): our choice of cutoff ensures that $\mt{S}(0) = S_{0|0,0}(0)$. 

\begin{figure}[ht!]
    \centering
    \includegraphics[width = 0.7\textwidth]{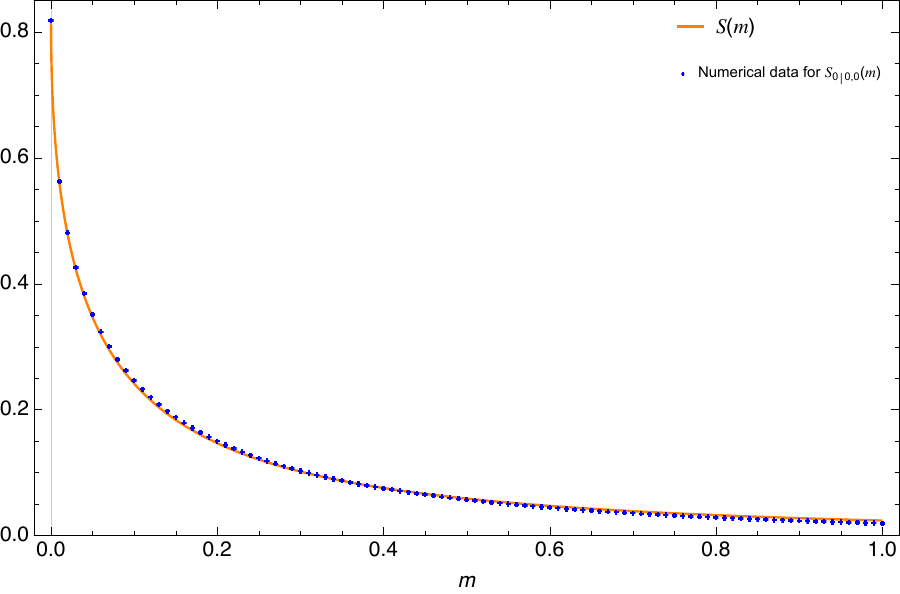}
    \caption{The approximate correction coefficient $\mt{S}(m)$ versus the numerical data. The masses are given in terms of units with $\frac{\bar{g}^2 N}{2\pi} = 1$}
    \label{Smass}
\end{figure}
The plot of $\mt{S}(m)$ versus the numerical data for $S_{0|0,0}(m)$ is given on figure \ref{Smass}. We see that the $S_{0|0,0}$ rapidly drops with an increase of $m$. Surprisingly enough, (\ref{SAppr}) provides a very good approximation despite its simplicity. Another interesting observation is that the other coefficients $S_{0|lk}$ drop at a much slower rate. For example,
\begin{equation}
    \left. \dfrac{S_{0|0,0}}{S_{0|0,1}}\right|_{m=1} = 7.0;\quad~\quad \left.\dfrac{S_{0|0,0}}{S_{0|0,1}}\right|_{m=0} = 64.
\end{equation}
It means that away from the chiral limit the contributions from the excited states of $B$ and $\pi$ become more relevant.

{Finally, it is interesting to compare our findings with the analysis of \cite{Trittmann:2024jkf}. While the latter was about the smallness of corrections in adjoint $\text{QCD}_2$ rather than in fundamental theory we consider, we can still point out some similarities. The main result of \cite{Trittmann:2024jkf} is the cancellation of certain leading ($1/N^2$ for adjoint matter) corrections to the spectrum. More specifically, only the terms that preserve the parton number were studied. A similar effect is manifest for the large $N$ single-meson states: the corrections change the coupling from $g^2$ to $\bar{g}^2 = g^2(1-1/N^2)$ in (\ref{LCHam}). The contributions $\sim 1/N$ to the quadratic term cancel each other and the four-meson parton-preserving term does not contribute due to normal ordering as we discussed before. On the other hand, we studied the corrections that arise from the parton number changing term. As it is mentioned in \cite{Trittmann:2024jkf}, such terms exist in adjoint QCD and also yield leading $\sim 1/N^2$ corrections while their smallness cannot be easily explained. Our results suggest that it is related to the mass gap in the adjoint theory.}

{More precisely, Fig. \ref{Smass} implies that the dependence of the correction on the meson mass mostly comes from the kinematic expression in the denominator of (\ref{GenCorr}) rather than the changes in the vertex function. Namely, the expression (\ref{SAppr}) was obtained assuming that the vertex function stays the same as at $m = 0$ (and using the cutoff approximation) while the pion mass can be nonzero. Effectively, it shows what would happen if the ground state pion was not massless but the vertex remained the same. Nevertheless, it is in a good agreement with the actual values of corrections. For large $\mu_0$ we get $\mt{S} \sim \frac{\mt{S}^3_0}{3\mu_0^2}$. Hence, if in the adjoint case $\mt{S}_0 \lesssim 1$ as we had for fundamental matter, the correction indeed should be small as $\mu_0^2 \sim 10$ \cite{Dempsey:2022uie}. Thus, our finding complement ones of \cite{Trittmann:2024jkf}, giving a possible explanation why the corrections from parton number violating terms are small.}

In summary, while $S_{0|0,0}$ from (\ref{SSing}) is indeed finite in the chiral limit, it still drastically increases and for small $N$ the corresponding correction becomes comparable with the large $N$ result $\xi_0$ from (\ref{ThEqHl}). This is closely tied to $\pi_0$ becoming massless, so we can conjecture that gapless theories would receive substantial $1/N$ corrections while gapped ones would not. Nevertheless, the qualitative structure of the spectrum remains the same as in the large $N$ limit. In the next section we will use a different numerical approach to check these results.

\subsection{The lightest meson at small $m$}\label{LightMes}
To finish this section, let us discuss the small $m$ limit of an even simpler example~--- $N_f = 1$ theory. The spectrum for $N=3$ and $m=2^{-10}$ was studied in \cite{Anand:2021qnd} with Lighcone Conformal Truncation (LCT). For instance, the lightest state mass squared is $\mu_{\text{num}}^2 = 0.003125$. On the other hand, the 't Hooft model prediction is $\mu_{\text{tH}}^2 = 0.003343$\footnote{The coupling parameter set to 1 in \cite{Anand:2021qnd} is $\frac{g^2 N}{2\pi}$ rather than $\frac{g^2 N}{2\pi} \lb 1-\frac{1}{N^2} \rb$ (see the first two subsections of Section \ref{LQuant}), so for better precision we use $2^{-10} / \sqrt{1-1/N^2} $ as quark mass and multiply the meson mass squared obtained from 't Hooft equation by $1-1/N^2$.}. We see that these results are rather close. Moreover, there is a separate calculation which includes only states with maximal parton number $n_{\text{max}} = 4$, e.g. at most 2-meson. This is precisely the Hilbert space part responsible for the type of corrections we considered above, when the intermediate states have 2 mesons. The result is $\mu_{\text{num},(4)}^2 = 0.003334$, which is extremely close to the 't Hooft model result. As we will see, it happens as the $\ket{\pi_0\pi_0}$ loop in perturbation theory vanishes due to parity constraints.

The leading corrections in one-flavor case can be found from (\ref{GenCorr}):
\begin{equation}
    \delta \mu_{0}^2 = -\dfrac{2}{N}\sum_{\substack{l,k\\
    (l+k) \mod 2 = 1}} \ili_0^1 dz \dfrac{\mt{T}^2(\pi_0|\pi_l\pi_k;z)}{\mu_l^2(1-z)+\mu_k^2 z - \mu_0^2 z(1-z)},
\end{equation}
where $\mu_n$ is the mass of $\pi_n$ as before. We used the parity relation (\ref{SymTEq}) to simplify the expression, as a consequence the intermediate state $\ket{\pi_l\pi_k}$ parity should be even (remember that the ground state is parity-odd). The vertex functions and corrections can be computed using numerical methods of Appendix \ref{HLVFunc}. Note that when $m\to 0$ the vertex function $\mt{T}(\pi_0|\pi_l\pi_k;z)$ should vanish as we discussed in Subsection \ref{VertProp}. From the explicit expression (\ref{TVertGen}) and the wave function endpoint behavior (\ref{PhiAsymp}) it follows that $\mt{T}(\pi_0|\pi_l\pi_k;z) \sim m$. Then $\delta \mu_0^2 \sim (m^2 \log m) /N \sim (\mu_0^4 \log \mu_0) /N$ (logarithmic part comes from either $k=0$ or $l=0$). The numerical computations give
\begin{equation}
    \delta \mu_0^2 = -\dfrac{23 \cdot 10^{-6}}{N} = -7.8\cdot 10^{-6}.
\end{equation}
LCT result for $n_{\text{max}} = 4$ is $-9.2 \cdot 10^{-6}$, which is rather close taking into account that we did not consider the higher order corrections which are also present.  Hence, our results are consistent with the numerical study.

The corrections with $n_{\text{max}} = 4$ become much stronger if $\ket{\pi_0 \pi_0}$ intermediate state is not forbidden by symmetries. It happens when $N_f > 1$. Let us consider the $N_f = 2$ theory we studied above with $M=m \to 0$. Then the correction to the ground state $B$-meson mass coming from the lightest meson loop is
\begin{equation}
    \delta \mu_{B_0|0,0}^2 = -\dfrac{2}{N} \ili_0^1 dz \dfrac{\mt{T}^2(\pi_0|\pi_0\pi_0;z)}{\mu_0^2[1-z(1-z)]}.\label{LeadCorrChir}
\end{equation}
There is an extra factor of $2$ as there are two possibilities for the intermediate states: $R_1 = B$, $R_2 = \pi$ and $R_1 = \eta$, $R_2 = B$. Due to small $m$ scaling of $\mu_0$ (\ref{MuAsymp}) we get $\delta \mu_{B_0|0,0}^2 \sim m/N \sim \mu_0^2/N$. Hence at small $m$ this correction is parametrically larger than other contributions with $l\neq 0$ or $k\neq 0$. Using the approximation for $\mt{T}(\pi_0|\pi_0\pi_0;z)$ at $m\to 0$ from Appendix \ref{EqMass} we find:
\begin{equation}
    \delta \mu_{B_0|0,0}^2 = -0.559\,\dfrac{m}{N}.\label{SmallMCorr}
\end{equation}
For $m=2^{-10} \sqrt{1-1/N^2}$ and $N=3$ it gives $-170 \cdot 10^{-6}$, so for a 2-flavor model the corrections from $n_{\text{max}} = 4$ would be much more noticeable.

So far we observed that the corrections behave smoothly at $m\to 0$. However, there can be issues at higher orders of perturbation theory. For example, consider the 4-order ($\sim 1/N^2$) correction to $\mu_{B_0}$ in the equal mass $N_f = 2$ theory. In quantum mechanics with a perturbation potential $V$ and basis of states $\ket{n}$ the 4-th order perturbative correction to the energy $E_n$ is as follows, provided $\bra{n}V\ket{n} = 0$:
\begin{equation}
    E_n^{(4)} = \sum_{k_1,k_2,k_3} \dfrac{V_{nk_3} V_{k_3k_2} V_{k_2 k_1} V_{k_1 n}}{E_{nk_1} E_{nk_2} E_{nk_3}} - \sum_{k_1,k_2} \dfrac{|V_{nk_2}|^2 |V_{nk_1}|^2}{E_{n k_2}^2 E_{n k_1}}.
\end{equation}
Here $V_{ml}$ are matrix elements and $E_{ml} = E_{m}-E_l$. In our case $\ket{n} = \ket{B_0}$, let us assume $\ket{k_1} = \ket{B_0 \pi_0}$, $\ket{k_2} = \ket{B_2}$, $\ket{k_3} = \ket{B_0 \pi_0}$ in the first term. We use $\ket{B_2}$ and not, say, $\ket{B_1}$ so that vertex integrals would not be zero from $z\to 1-z$ reflection properties. Then its denominator $\sim \mu_0^4 \sim m^2$ from $E_{n k_1} E_{n k_3}$. On the other hand, in the numerator only $V_{k_1 n}$ and $V_{n k_3}$ are small as $\mt{T}(\pi_2|\pi_0\pi_0;z)$ does not vanish in $m\to 0$ limit. Hence, the numerator $\sim m^2$ and this term is $m$-independent at small $m$. It means that the correction becomes parametrically larger than $\mu_0^2 \sim m$ when $m\to 0$. This cannot be compensated by the second term, as it at most $\sim m$. At higher orders of perturbation theory the corrections even become divergent at small $m$ as we can insert more $\ket{B_0 \pi_0}$ intermediate states: each gives an extra $1/m$ factor. In Section \ref{ThrSt} dedicated to threshold states we will explain how to properly use degenerate perturbation theory in such situations. We will show that the corrections from such intermediate states do not diverge, but rather $\sim m/N$ when $m \ll 1/N$ and the infinite $N$ spectrum does not change its structure. So at small $m$ some of the $1/N^2$ corrections turn into $1/N$ and $\delta \mu_{{B_0}|0,0}^2$ is not the only correction $\sim m/N$.

When $N_f = 1$ there are similar issues at higher orders of perturbation theory which they show up only for $n_{\text{max}} \ge 6$ due to parity-related cancellations. This explains why the corrections from $n_{\text{max}} = 6$ states found by LCT are rather substantial and much larger than from $n_{\text{max}} = 4$ states. Normally they are $\sim 1/N^2$, but they could be amplified to $1/N$ at small $m$.

Finally, at small quark mass bosonization should provide a good description of low energy effective theory \cite{Witten:1983ar,Steinhardt:1980ry,Frishman:1992mr}. In the 1-flavor case it should be sine-Gordon theory with interaction Largrangian
\begin{equation}
    \mt{L}_{\text{int}} = \mu'^2 \cos \dfrac{2\sqrt{\pi}}{\sqrt{N}} \phi.
\end{equation}
The mass parameter $\mu'$ is expressed in terms of quark mass:
\begin{equation}
    \mu' = \sqrt{2}\left[ \dfrac{N^{1-\Delta_C/2} e^{\gamma} m }{2}\right]^{\frac{1}{1+\Delta_C}},\quad \Delta_C = \dfrac{N-1}{N};~\dfrac{g^2 N}{2\pi} = 1.
\end{equation}
Then sine-Gordon solitons (kinks) correspond to baryons, and mesons are kink-antikink bound states. The baryon mass $\mu_B$ can be obtained semiclassically \cite{Dashen:1975hd}: $\mu_B = \frac{2\mu'}{\sqrt{\pi N}}(2N-1)$. The lightest meson mass is then expressed in terms of $\mu_B$:
\begin{equation}
    \mu_0 = 2\mu_B \sin \dfrac{\pi}{2(2N-1)}.
\end{equation}
For our parameters it gives $\mu_0^3 = 0.00415$, which is clearly significantly different from LCT and 't Hooft model results. Moreover, at large $N$ it scales as
\begin{equation}
    \mu_0^2 = \dfrac{4 \pi \mu'^2}{N} = \dfrac{4\pi m e^\gamma}{\sqrt{N}},
\end{equation}
so bosonization prediction for $\pi_0$ mass is parametrically lower than 't Hooft model one. For that to be the case, the pion mass should receive corrections that do not drop with $N$. We did not observe such effects in perturbation theory, even though at higher orders one has to use degenerate theory to avoid divergences.

The fact that the small $m$ limit is described by sine-Gordon theory, although with mass parameter different from $\mu'$, is supported by \cite{Anand:2021qnd}. The ratios between bound state masses are $\mu'$ independent. In particular, the theory predicts 2-meson bound states with mass $\mu_{2M}$ slightly below the threshold $2\mu_0$ \cite{Dashen:1975hd}:
\begin{equation}
    \dfrac{\mu_{2M}}{2\mu_0} = \dfrac{\sin \dfrac{\pi}{2N-1}}{\sin \dfrac{\pi}{2(2N-1)}} = 1-\dfrac{\pi^2}{32 N^2} + O(N^{-3}),\quad N\to \infty.
\end{equation}
The numerics of \cite{Anand:2021qnd} confirmed this ratio. Note that the large $N$ expansion for $\mu_{2M} - 2\mu_0$ looks like a standard perturbative series with integer powers of $N$. On the other hand, in Section \ref{ThrSt} we will study a different kind of near threshold states, for which the difference between the bound state mass and the threshold is $\sim 1/N^{2/3}$. This happens because the usual perturbative expansion breaks down.

One possible way to fix the discrepancy between bosonization and 't Hooft model is to redefine the interaction term at a different scale. Namely, to the interaction should be normal-ordered, and to do that one needs to introduce a scale of normal-ordering \cite{Coleman:1974bu}. The connection formula for normal orderings at different scales $m'$ and $\Lambda$ is as follows:
\begin{equation}
    N_{\Lambda'} \cos \beta\phi = \lb\dfrac{\Lambda}{\Lambda'} \rb^{\frac{\beta^2}{4\pi}} N_\Lambda \cos\beta\phi.
\end{equation}
Here $N_\Lambda$ is normal ordering at scale $\Lambda$, the Lagrangian above is normal-ordered at $m'$. To make the large $N$ meson mass reproduce the 't Hooft model result (\ref{MuAsymp}), we need to replace $\mu'^2$ with $\tilde{\mu}^2 = \frac{N m}{2\sqrt{3}}$. Hence, we need to renormal-order at a scale
\begin{equation}
    \Lambda = \mu' \lb \dfrac{\tilde{\mu}}{\mu'} \rb^{2N} \sim N^{\frac{2N+1}{4}}.
\end{equation}
It looks very unnatural that one has to renormal-order to such a large scale, so there should be a different explanation why bosonization gives incorrect scaling.

\section{Discretized approach}\label{DLCQ}
Now we will apply the DLCQ approach to repeat the calculations from the previous section in order to test them and see how well this method fares. As we mentioned in the introduction, the idea is to discretize the momentum space to make the space of states finite-dimensional. The original method does not work well for small but nonzero quark masses \cite{Hornbostel:1988fb,Anand:2021qnd}: the meson masses converge as $K^{-2\beta}$, where $K$ is discretization size and $\beta$ is the power defining wave function boundary behavior (\ref{BetEq}) for the light quark mass $m$. It happens because meson wave functions vary rapidly near endpoitns. As $\beta \sim m$, this requires using very high $K$ for $m < 1$ to achieve good precision, which leads to very high Hilbert space dimension. A way to avoid this and make convergence $1/K$ was proposed in \cite{vandeSande:1996mc} and tested for the 't Hooft equation. Here we will use it to compute the corrections and see if it could be trusted for finite $N$.

\subsection{Lightcone discretization}
In order to discretize the lightcone momentum space we compactify the lightcone spatial coordinate $x^-$: $x^- \sim x^- + \pi R$. We will assume antiperiodic boundary conditions for the fermions so the momenta take values $p = \frac{n}{R}$ with odd $n$. This choice helps to avoid the fermionic zero mode issue. It leads to the following correspondence rules for the lighcone Hamiltonian (\ref{LcHam}):
\begin{equation}
    \ili dk \to \dfrac{2}{R} \sum_n,\quad \delta(k-k') \to \dfrac{R}{2} \delta_{nn'},\quad M^0_{kk',ii'} \to \dfrac{R}{2} M^0_{nn',ii'}.
\end{equation}
We will diagonalize the Hamiltonian (\ref{LcHam}) with $\frac{\bar{g}^2 N}{2\pi} = 1$ in the space of states with fixed momentum $\frac{K}{R}$. The mass spectrum is given by the eigenvalues of $\mtf{m}^2=2 P^+ P^-$ as before.

First, let us diagonalize the infinite $N$ Hamiltonian, then we will be able to use the eigenstates in perturbation theory to find the corrections. It is possible to study the full theory with DLCQ as well as the number of states stays finite, but it grows exponentially if we include multi-particle states. The large $N$ Hamiltonian is quadratic in $M^0$ and does not mix flavors, so we can consider only single-particle states (a particle in this context is a state created by $M^{0\dagger}$). 

Before moving to numerical computations, we need to take the small quark mass behavior into account. Following \cite{vandeSande:1996mc}, we modify the integrand of the second term of (\ref{LcHam}) as follows:
\begin{equation}
\begin{aligned}
    &~M^{0\dagger}_{kk',ii'} M^0_{pp',ii'} - M^{0\dagger}_{kk',ii'}M^0_{kk',ii'} \to\\
    &~\left(M^{0\dagger}_{kk',ii'} M^0_{pp',ii'} - M^{0\dagger}_{kk',ii'}M^0_{kk',ii'}\dfrac{\gamma_i y^{\beta_i}(1-y)^{1-\beta_i} + \gamma_{i'} y^{1-\beta_{i'}}(1-y)^{\beta_{i'}} }{\gamma_i x^{\beta_i}(1-x)^{1-\beta_i} + \gamma_{i'} x^{1-\beta_{i'}}(1-x)^{\beta_{i'}}}\right)\\
    &+M^{0\dagger}_{kk',ii'}M^0_{kk',ii'}\left(\dfrac{\gamma_i y^{\beta_i}(1-y_i)^{1-\beta_i} + \gamma_{i'} y^{1-\beta_{i'}}(1-y)^{\beta_{i'}} }{\gamma_i x^{\beta_i}(1-x_i)^{1-\beta_i} + \gamma_{i'} x^{1-\beta_{i'}}(1-x)^{\beta_{i'}}}-1 \right),\quad x = \dfrac{k}{k+k'},~y=\dfrac{p}{p+p'}.
\end{aligned}
\end{equation}
Here $\gamma_i = 1$ if $0<m_i<1$ and is 0 otherwise~--- for $m_i>1$ or for $m_i=0$ the meson wave functions change slowly so we do not need extra terms. Note that we use a different boundary behavior term compared to \cite{vandeSande:1996mc}: for our choice the integrals over $p'$ are expressed in terms of elementary functions. We also do not take into account the boundary behavior of $1/N$ suppressed terms: as we will see, the precision is good enough even without it. After discretization, the matrix elements between $\ket{nn',ii'} = M^{0\dagger}_{nn',ii'}\ket{0}$ are as follows:
\begin{equation}
\begin{aligned}
    \bra{ll',ii'} 2P^+P^- \ket{nn',ii'} &= \left[K\lb\dfrac{m_i^2}{n} + \dfrac{m_{i'}^2}{n'} \rb \right.\\
    &- \gamma_i\gamma_{i'}\dfrac{K^2}{nn'}  - K\gamma_{i}\dfrac{\pi \cot(\pi\beta_i)(n-\beta_i K)n^{\beta_i-1}n'^{-\beta_i} - \pi \csc(\pi\beta_i)}{\gamma_i n^{\beta_i} n'^{1-\beta_{i}}+ \gamma_{i'}n^{1-\beta_{i'}} n'^{\beta_{i'}}}\\
    &-\left.K\gamma_{i'}\dfrac{\pi \cot(\pi\beta_{i'})(n'-\beta_{i'} K)n^{-\beta_{i'}}n'^{\beta_{i'}-1} - \pi \csc(\pi\beta_{i'})}{\gamma_i n^{\beta_i} n'^{1-\beta_{i}}+ \gamma_{i'}n^{1-\beta_{i'}} n'^{\beta_{i'}}}\right]\delta_{nl}\delta_{n'l'}\\
    &-2 K\lb \dfrac{1}{(n-l)^2} - \delta_{n,l}\sum_{k < K} \dfrac{1}{(k-n)^2} \dfrac{\gamma_i k^{\beta_i} k'^{1-\beta_{i}}+ \gamma_{i'}k^{1-\beta_{i'}} k'^{\beta_{i'}}}{\gamma_i n^{\beta_i} n'^{1-\beta_{i}}+ \gamma_{i'}n^{1-\beta_{i'}} n'^{\beta_{i'}}} \rb,
\end{aligned}
\end{equation}
here $n+n' = l+l' =k+k'= K$. Also, the factor at $\frac{1}{(k-n)^2}$ is $1$ when $\gamma_i = \gamma_{i'}=0$. The powers $\beta_i$ are defined by (\ref{BetEq}) as before. The infinite size (continuum in momentum space) limit is then achieved by $K\to \infty$. We also remove the singular terms from the above expression by hand: it amounts to removing the gauge field zero mode as we explained in Section \ref{LcQt}.

It is important to understand the limitations of this approach. Note that this way of diagonalization is essentially a discretization of the 't Hooft equation (\ref{Th-Eq}) with a step size of $\frac{2}{K}$. It means that this approach will fail if the wave function significantly changes at such scale. While we removed this effect at small quark mass by using the endpoint behavior, it happens in the case of heavy quarks. In Subsection \ref{HLlimit} we observed that the heavy-light wave function is mostly concentrated in the region of size $1/M$ ($M$ is the heavy quark mass). Hence one cannot properly take the heavy quark limit while keeping $K$ finite and for large $M$ we need to make $K$ at least comparable to $M$.

\begin{figure}[ht]
\centering
\begin{minipage}[h]{0.49\linewidth}
\center{\includegraphics[width=1\textwidth]{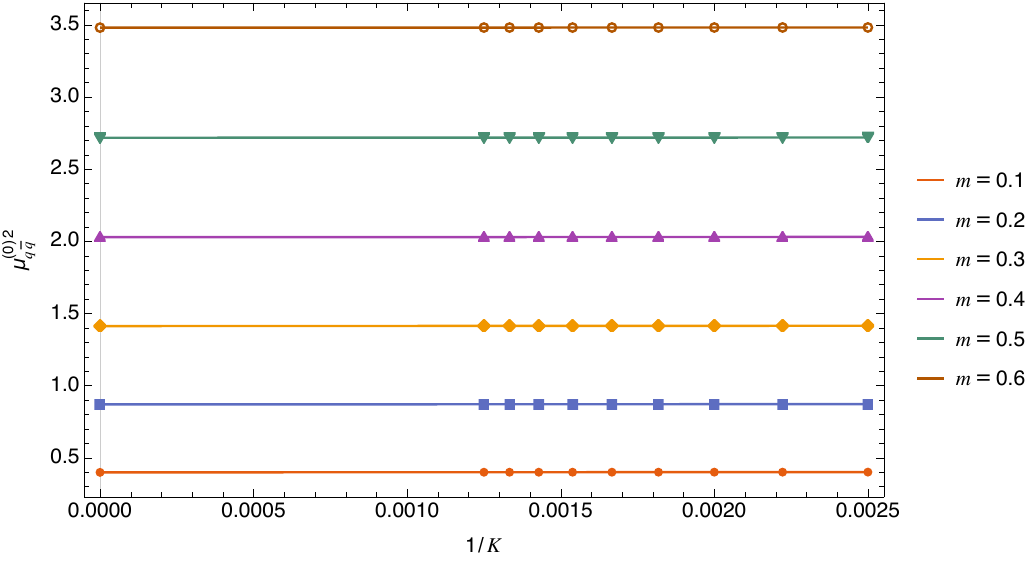} (a)}
\end{minipage}
\begin{minipage}[h]{0.49\linewidth}
\center{\includegraphics[width=1\textwidth]{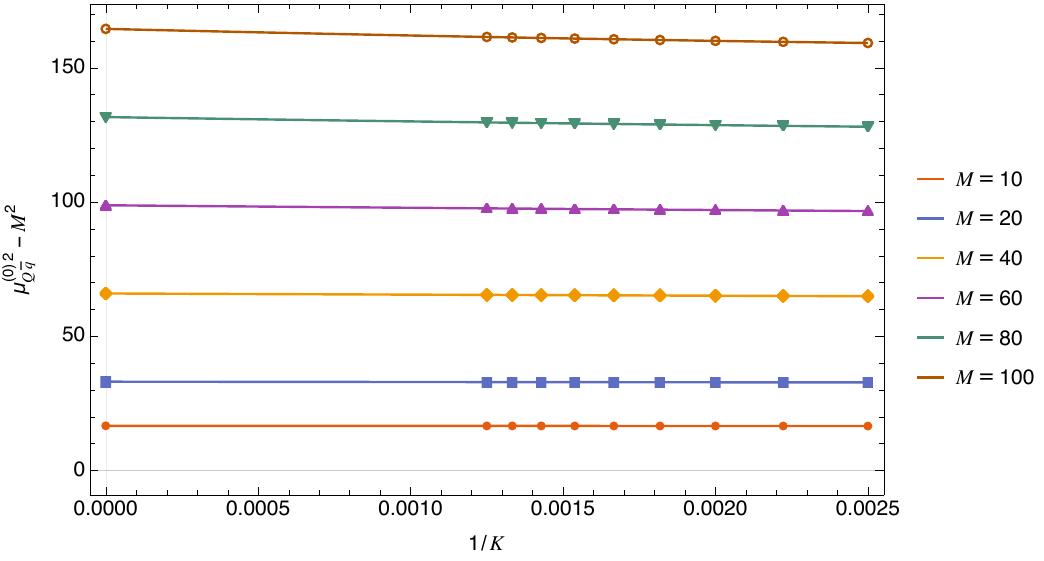} (b)}
\end{minipage}
\hfill

\begin{minipage}[h]{0.49\linewidth}
\center{\includegraphics[width=1\textwidth]{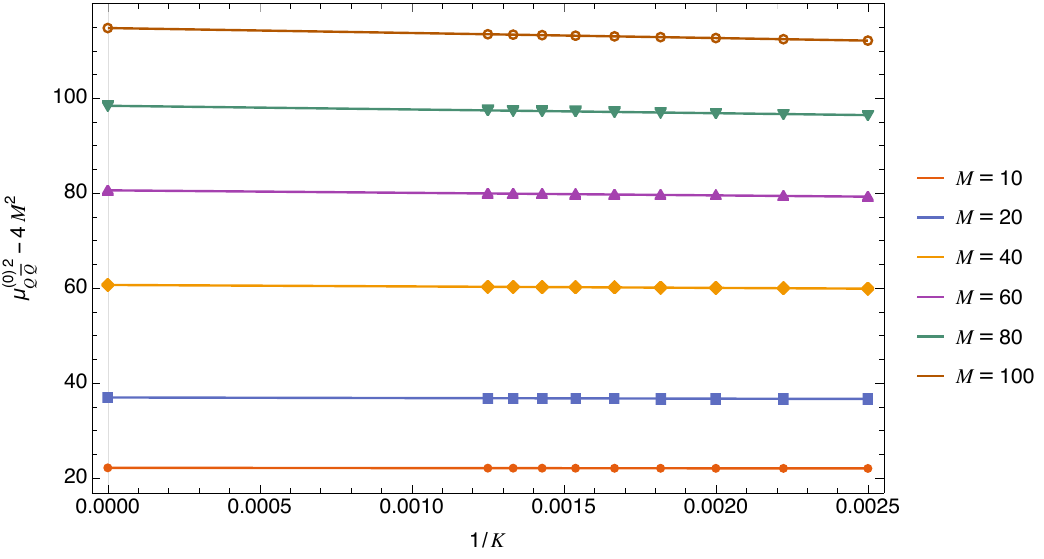} (c)}
\end{minipage}
\begin{minipage}[h]{0.49\linewidth}
\center{\includegraphics[width=1\textwidth]{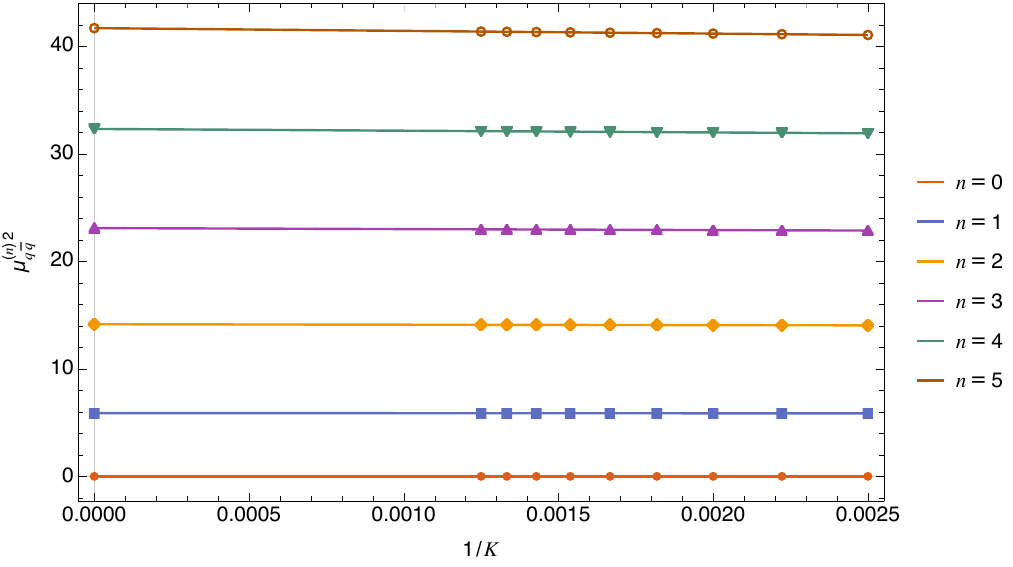} (d)}
\end{minipage}
\caption{The infinite $N$ DLCQ calculations compared to the continuum limit: a) The ground state pion mass squared for quark mass $m$; b) The dynamical part of the ground state heavy-light meson mass squared for zero light quark mass and heavy quark mass $M$; c) The dynamical part of the ground state heavy-heavy meson for quark mass $M$; d) The squared masses of the $n$-th pion excited state for the massless quark case $m=0$. The solid lines represent the cubic extrapolation of the DLCQ data to $K=\infty$, the points at infinite $K$ correspond to 't Hooft equation solutions. The masses are given in terms of units with $\frac{\bar{g}^2 N}{2\pi} = 1$}\label{DLCQ_Inf}
\end{figure}
The DLCQ numerical results for different mesons are presented on Figure \ref{DLCQ_Inf}, the maximal momentum number in our computations is $K=800$. The infinite $K$ masses used for comparison were obtained using the numerical approach described in Appendix \ref{MassThEq}. We again consider two quark flavors: a heavy $Q$ with mass $M$ and a light quark $q$ with mass $m$. We see that the DLCQ results for $B$ (b) and $\eta$ (c) mesons are in good agreement with the 't Hooft equation even for $M = 100$. The same holds for the pion ($q\bar{q}$) states with $m=0$ (a) and nonzero but small $m$ (d).

{Let us also clarify some points regarding the basis size. First, at finite $N$ reaching $K=800$ is impossible: the number of states grows very fast. As discussed in \cite{luch20241+}, the generating function for the total number of states at fixed $K$ and infinite $N$ is related to the number of plane partitions. For instance, at $K=800$ one gets the number of states $\sim 10^{45}$, which is computationally infeasible. On the other hand, at $N\to \infty$ the particle production is suppressed, and in this subsection we used only single-meson states  for the basis, their number at $K=800$ is 400. In the next subsection we will use the infinite $N$ basis to check if $1/N$ corrections we found before are reproduced. To do this it is enough to add only 2-meson states\footnote{The truncation in meson number is rather efficient even at small $N$ as evidenced by \cite{Anand:2021qnd}}. Note that the flavor charged is conserved, so if we study $Q\bar{q}$ mesons, the two-meson states should have the form $(Q\bar{Q})(Q\bar{q})$ or $(q\bar{q})(Q\bar{q})$. In general, the number of such states scales as $K^3$, which still can be rather large. We will further reduce the number of states by truncating in the quantum numbers of the infinite $N$ 2-meson states. The continuum results imply that we only need the first few terms of such truncation to have good precision. Then for fixed quantum numbers the total number of such states with momentum $K$ is $\sim K$ and reaching even $K=800$ is possible.}

\subsection{Corrections in DLCQ}
\begin{figure}[!hb]
\centering
\begin{minipage}[h]{0.435\linewidth}
\center{\includegraphics[width=1\textwidth]{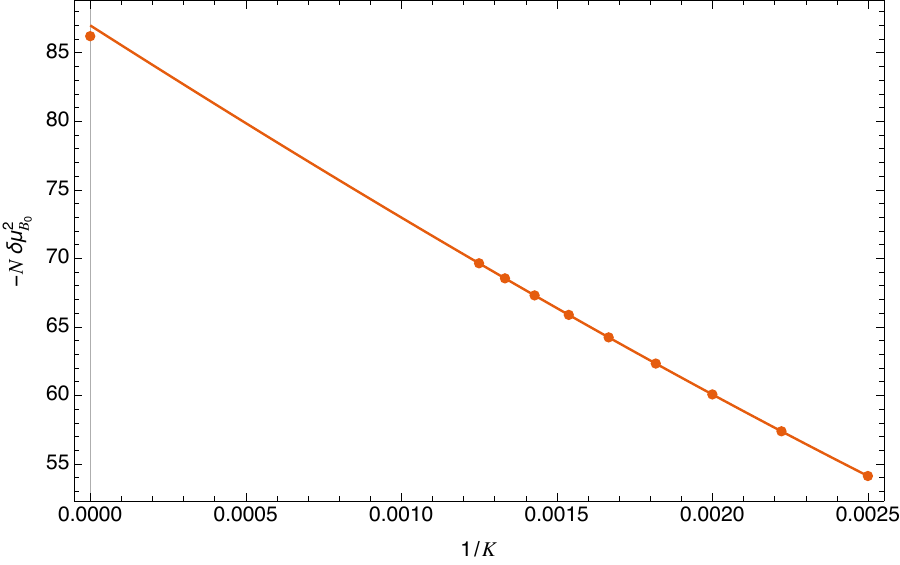} (a)}
\end{minipage}
\begin{minipage}[h]{0.49\linewidth}
\center{\includegraphics[width=1\textwidth]{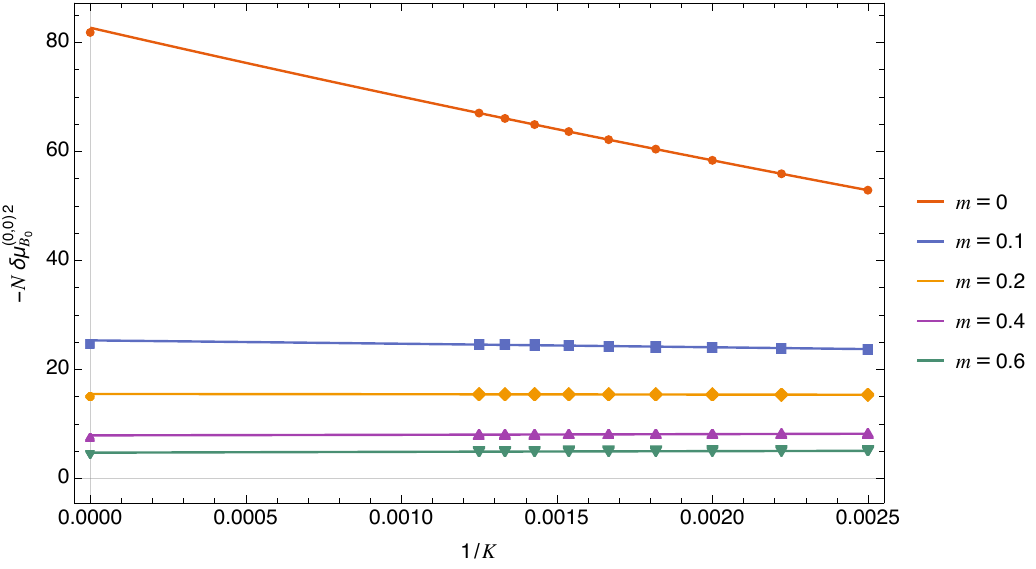} (b)}
\end{minipage}
\hfill
\caption{The DLCQ result for $1/N$-corrections: a) Part of the correction (\ref{DLCQ_Corr}) associated with $B\pi$ loop with $I,J \le 4$ to compare with the continuum calculation, the extrapolation gives $N\delta \mu^2_{B_0} = 87.0$. The heavy quark mass is $M=100$, the light quark is massless. The point at infinite $K$ represents the heavy-light limit continuum result (\ref{DelMuBCont}). b) The correction from the ground state part $\ket{B_0\pi_0}$ for the light quark mass $m$ with still fixed $M=100$. The infinite $K$ points correspond to the data from Fig. \ref{Smass}. The masses are given in terms of units with $\frac{\bar{g}^2 N}{2\pi} = 1$}\label{DLCQ_Qq}
\end{figure}
The $1/N$-corrections to the masses can be found from the standard perturbation theory for $2P^+P^-$. The leading interaction term $U^{(1)}=2\sqrt{N}P^+P^{-(1)}$ in the two-flavor case has the following form:
\begin{equation}
    U^{(1)} = 2K\sum_{i,i',j;n,n',l,l',o}\left[ \dfrac{M^{0\dagger}_{no,ij}M^{0\dagger}_{ll',ji'}M^0_{nn',ii'}}{(l+o)^2} \delta_{o+l+l',n'} - \dfrac{M^{0\dagger}_{ll',ij}M^{0\dagger}_{on',ji'}M^0_{nn',ii'}}{(n-l)^2} \delta_{o+l+l',n} + \text{h.c.}\right],
\end{equation}
here $n+n' = K$. We need to use the second-order perturbation theory, we will study the case $B$-meson case $i=Q$, $j=q$. Let us denote the eigenstates as follows:
\begin{equation}
\begin{gathered}
    \ket{\pi_I(n)} = \sum_{k<n} C_{qq,k}^{(I)}(n) M^{0\dagger}_{k,n-k;qq}\ket{0},\quad\ket{B_I(n)} = \sum_{k<n} C_{Qq,k}^{(I)}(n) M^{0\dagger}_{k,n-k;Qq}\ket{0},\\ \ket{\eta_{I}(n)} = \sum_{k<n} C_{QQ,k}^{(I)}(n) M^{0\dagger}_{k,n-k;QQ}\ket{0}.
\end{gathered}
\end{equation}
We will also denote the corresponding masses by $\mu^{(I)}_{qq}(n)$, $\mu^{(I)}_{Qq}(n)$ and $\mu^{(I)}_{QQ}(n)$. Note that the number of 1-particle states is bounded: $I \leq \frac{n}{2}-1$ (we count the states starting with $I=0$). The two-particle states like $\ket{B_I(n)\pi_J(K-n)}$ are simply given as products of such operator expressions. 

The correction to the $B_0$ mass squared is then given by the following expressions:
\begin{equation}
\begin{aligned}
    N\delta \mu^2_{B_0} &= -\sum_{I,J,n} \dfrac{\left|\bra{B_I(n)\pi_J(K-n)}U^{(1)}\ket{B_0(K)}\right|^2}{K\lb \dfrac{\mu^{(I)2}_{Qq}(n)}{n} + \dfrac{\mu^{(J)2}_{qq}(K-n)}{K-n}\rb-\mu_{Qq}^{(0)2}(K)} \\
    &-\sum_{I,J,n} \dfrac{\left|\bra{\eta_I(n)B_J(K-n)}U^{(1)}\ket{B_0(K)}\right|^2}{K\lb \dfrac{\mu^{(I)2}_{QQ}(n)}{n} + \dfrac{\mu^{(J)2}_{Qq}(K-n)}{K-n}\rb-\mu_{Qq}^{(0)2}(K)}.\label{DLCQ_Corr}
\end{aligned}
\end{equation}
Here $n$ is even and $1<n<K$. The relevant matrix elements can then be easily computed:
\begin{equation}
    \bra{B_I(n)\pi_J(K-n)}U^{(1)}\ket{B_0(K)} = 2K \sum_{s,t} \dfrac{C^{(I)}_{Qq,s}(n) C_{qq,t}^{(J)}(K-n) [C^{(0)}_{Qq,s}(K)-C^{(0)}_{Qq,n+t}(K)]}{(n-s+t)^2},
\end{equation}
where $1\le s \le n-1$, $1\le t\le K-n-1$. In the same manner,
\begin{equation}
    \bra{\eta_I(n)B_J(K-n)}U^{(1)}\ket{B_0(K)} = 2K \sum_{s,t} \dfrac{C^{(I)}_{QQ,s}(n) C_{Qq,t}^{(J)}(K-n) [C^{(0)}_{Qq,s}(K)-C^{(0)}_{Qq,n+t}(K)]}{(n-s+t)^2}.
\end{equation}

Now it is very straightforward to compute the corrections. The results are presented on Figure \ref{DLCQ_Qq}. As in the subsection \ref{NumResSec} we sum over the first 25 two-particle states $\ket{B_I\pi_J}$: $I,J\le 4$ (a). The heavy-light continuum approach given by (\ref{SCorr}), (\ref{DelMuBCont}) predicts $-N\delta \mu_{B_0}^2 = MS_{0,0} = 86.2$, which is close to the extrapolated DLCQ value $87.0$. The difference is of order $\frac{1}{M}$ so it is within the range of $\frac{1}{M}$ corrections to the heavy-light limit. DLCQ allows for faster calculations and the inclusion of the higher excited two-particle states slightly modifies the answer yielding $-N\delta \mu_{B_0}^2 = 87.8$. We also see that DLCQ works well for small nonzero $m$ (b), even though we did not take into account the boundary behavior of the three-point vertex. This is because the $B_0\pi_0$ loop correction's dependence on $m$ mostly comes from $\pi_0$ mass as the approximation (\ref{SAppr}) from the previous subsection works very well. We did not include the corrections coming from $B\eta$-loops as they are heavily suppressed (the related correction is $\sim 10^{-6}$).

{One issue that might arise at finite $N$ is that some states in the infinite $N$ basis become null \cite{Dempsey:2022uie}. The finite $N$ mesonic basis is generated by the discretized version of (\ref{CompOp}):
\begin{equation}
M^\dagger_{nn',ii'} = \dfrac{1}{\sqrt{N}}b^{\alpha\dagger}_{n,i}d^{\alpha\dagger}_{n',i'},\quad \ket{(n_1n_1',i_1i_1')\dots (n_\mt{N} n_\mt{N}',i_\mt{N}i_\mt{N}')}= M^\dagger_{n_1 n_1',i_1i_1'}\dots M^\dagger_{n_\mt{N} n_\mt{N}',i_\mt{N}i_\mt{N}'} \ket{0}.
\end{equation}
Here we also presented a typical basis state, $\mt{N}$ is the meson number. In our analysis we used the canonical $M^{0\dagger}$ operators from (\ref{OpExp}) rather than physical $M^\dagger$. In the leading order in $1/N$ we simply need to replace $M^{0\dagger}\to M^\dagger$ to switch to the physical basis. However, due to the fermionic nature of $b^\dagger$ and $d^\dagger$, the states given by different sets $(n_1n_1',i_1i_1')\dots (n_\mt{N} n_\mt{N}',i_\mt{N}i_\mt{N}')$ are not linearly independent: the basis is overdetermined. The respective constraints are known as trace relations \cite{Dempsey:2022uie}. The simplest example is the case when a state has at least $N+1$ of $b^\dagger$ or $d^\dagger$ operators with the same flavor and momentum: it is null due to the exclusion principle. More generally, all relations were listed in \cite{luch20241+}, they have the following form:
\begin{equation}
    \sum_{\sigma \in S_{N+1}} M^\dagger_{n_{\sigma_1}n'_1,ii_1'}\dots M^\dagger_{n_{\sigma_{N+1}}n'_{N+1},i i_{N+1}'} = 0,\quad \sum_{\sigma \in S_{N+1}} M^\dagger_{n_1 n'_{\sigma_1},i_1i'}\dots M^\dagger_{n_{N+1} n'_{\sigma_{N+1}},i_{N+1}i'}=0.
\end{equation}
Here $S_{N+1}$ is the permutation group, the operators that participate in a relation should have the same flavor. The most important part for us is that the trace relations appear only for states with meson number $\mt{N}> N$. On the other hand, the states that we considered so far and will consider in the next sections are at most 2-meson. Hence, they are not affected by trace relations for any $N$ and survive all the way to $N=2$.}

To sum up, we see that DLCQ gives the same results for corrections as the continuum approach in the chiral limit and for sufficiently large light quark mass. The improved method of \cite{vandeSande:1996mc} gives good results for small $m$ as well even without any endpoint modifications to the three-point vertex. So, this method can be useful to study the full theory at small $m$.

\section{Effective theory}\label{EFT}
In this section we investigate the $\pi BB$ coupling from the effective theory point of view. The pion is represented by a real scalar field $\pi(x)$ while $B$ is a complex scalar. We know that in the chiral limit the massless pion should have derivative coupling and is parity-odd. Hence the simplest low energy effective Lagrangian for $B$, $\bar{B}$ and $\pi$ is as follows \cite{Grinstein:1994nx,Grinstein:1994hy}:
\begin{equation}
    \mt{L} = |\p_\mu B|^2 - \mu^2 |B|^2 + \dfrac{1}{2} \left[ (\p_\mu \pi)^2 - \mu_0^2 \pi^2 \right] - i \kappa \lc^{\mu\nu} \p_\mu \bar{B} \p_\nu B \pi. \label{EffLagr}
\end{equation}
Here $\mu$ is the $B_0$ mass, $\mu_0$ is the $\pi_0$ mass which we added for generality and the antisymmetric symbol is defined by $\lc_{01} = 1$. In the lightcone coordinates $x^{\pm}$ we get $\lc^{+-} = 1.$ 
\subsection{Lightcone quantization}\label{LcQt}
In order to find the dimensionless coupling $\kappa$ we need to compare the interaction term in (\ref{EffLagr}) with the exact vertex function (\ref{TVertGen}) when $z\to 1$ (low energy limit). The latter is given in the lightcone-quantized theory, so we will use lightcone interaction picture of (\ref{EffLagr}) for this comparison. The free field time-independent quantization is as follows \cite{Liu:1993, Heinzl:2000ht}:
\begin{equation}
\begin{aligned}
    B(x) &= \ili_0^{+\infty} \dfrac{d P^+}{2\sqrt{\pi P^+}} \left[ a_{P^+} e^{-i P^+ x^-} + b_{P^+}^\dg e^{i P^+ x^-}  \right];\\
    \pi(x) &= \ili_0^{+\infty} \dfrac{d P^+}{2\sqrt{\pi P^+}} \left[ c_{P^+} e^{- i P^+ x^-} + \text{h.c.} \right].
    \end{aligned}\label{FieldQuant}
\end{equation}
The creation/annihilation operators satisfy the canonical commutation relations:
\begin{equation}
    [a_{P^+_1}, a^\dg_{P_2^+}] = [b_{P^+_1}, b^\dg_{P_2^+}] = [c_{P^+_1}, c^\dg_{P_2^+}] = \delta(P^+_1-P^+_2).
\end{equation}
Note that the system is constrained unlike in the standard equal time quantization case. Namely, the conjugate momenta for free fields $\mtf{p}_{\bar{B}} = \p_- B$, $\mtf{p}_{B} = \p_-\bar{B}$, $\mtf{p}_{\pi} = \p_-\pi$ are clearly not independent. It turns out that the commutation relations have an additional $\frac{1}{2}$ factor compared to the equal time case:
\begin{equation}
    [\pi(x^-),\p_- \pi(y^-)] = [B(x^-),\p_- \bar{B}(y^-)] = [\bar{B}(x^-),\p_- B(y^-)] = \dfrac{i}{2} \delta(x^- - y^-),\quad x^+=y^+.\label{CanComRel}
\end{equation}
It also implies that the field operators themselves have nontrivial commutation relations on a slice $x^+ = \text{const}$.

We cannot use the operator expansion (\ref{FieldQuant}) for a quantization of (\ref{EffLagr}). The issue is that the interaction term contains time derivatives, so the conjugate momenta of $\hat{\pi}_B$ and $\hat{\pi}_{\bar{B}}$ get modified compared to the free field case:
\begin{equation}
    \mtf{p}_{B} = \p_- \bar{B} + i \kappa \pi \p_-\bar{B},\quad \mtf{p}_{\bar{B}} = \p_- B - i \kappa \pi \p_-B.\label{BShift}
\end{equation}
It means that the commutation relations (\ref{CanComRel}) and as a consequence the expansion (\ref{FieldQuant}) are no longer valid. To resolve this issue perturbatively, we can shift the field $B$:
\begin{equation}
    B\to B + \dfrac{i\kappa }{2}\dfrac{1}{\p_-} (\p_- B \pi),\quad \bar{B}\to \bar{B} - \dfrac{i\kappa}{2}\dfrac{1}{\p_-}(\p_- \bar{B} \pi).
\end{equation}
After this shift the Lagrangian becomes as follows up to a total derivative:
\begin{equation}
    \mt{L} = \mt{L}_0 - \dfrac{i \kappa \mu^2}{2} \left[\bar{B} \dfrac{1}{\p_-}(\p_- B \pi)  -  \dfrac{1}{\p_-}(\p_- \bar{B} \pi) B\right] + O(\kappa^2),\label{EffLagrNew}
\end{equation}
where $\mt{L}_0$ is the free theory Lagrangian. Now the interaction term contains no time derivatives in the leading order in $\kappa$, so (\ref{FieldQuant}) holds up to $O(\kappa^2)$. The subleading interaction in the above Lagrangian still contains the $x^+$ derivatives, but it is linear in them just as the original Lagrangian. In principle, we can continue (\ref{BShift}) as a series in $\kappa$ to cancel such derivatives order by order at the expense of adding more non-local terms.

Let us focus on the leading interaction: we expect $\kappa$ to be small at large $N$ as the mesons are free particles in the $N\to \infty$ limit. We can apply the operator expansion (\ref{FieldQuant}) to the interaction potential, mostly interested in the term containing $a^\dagger c^\dagger a$ as it corresponds to the $B\to \pi B$ vertex we studied before.
We find:    
\begin{equation}
    V_I^{(a^\dagger c^\dagger a)} = \dfrac{i \kappa \mu^2}{4 \sqrt{\pi}} \ili_0^{+\infty} \dfrac{d p^+ d \tilde{p}^+ d q^+}{\sqrt{p^+ \tilde{p}^+ q^+}} \delta(p^+ - \tilde{p}^+-q^+) \dfrac{1-z^2}{2z} a^\dg_{\tilde{p}^+} c^\dg_{q^+}a_{p^+} ,\quad z = \dfrac{\tilde{p}^+}{p^+}.\label{IntVert}
\end{equation}
In the limit $z\to 1$ we get $\frac{1-z^2}{2z} \to 1-z$, comparing it with the original theory lightcone Hamiltonian (\ref{PEff}) and the $z\to 1$ behavior of the vertex function $\mt{T}$ (\ref{VertDer}) in the chiral limit yields
\begin{equation}
    \left.\kappa\right|_{\mu_0=0} = \dfrac{2 \sqrt{\pi}}{\sqrt{N}}.\label{LamConst}
\end{equation}
We used that the creation and annihilation operators are defined up to a phase rotation which does not change the commutation relations so we only need to compare the absolute values. It means that $\kappa$ is defined up to a sign as it should be real in the Lagrangian (\ref{EffLagr}). Note that this coupling is $\mu$-independent.

This result is rather simple, but we cannot use it in perturbation theory. For example, the one-loop correction to the $B$-meson mass computed from original covariant Lagrangian (\ref{EffLagr}) has a logarithmic ultraviolet (UV) divergence, which is not true in the full theory. In order to avoid that we need to make $\kappa$ momentum-dependent by matching the vertex functions from (\ref{PEff}) and (\ref{VertDer}) at arbitrary $z$. It is natural to expect that this will introduce a cutoff at the 't Hooft coupling scale which we set to 1. Let us denote $p\wedge \tilde{p} = \lc_{\mu\nu} p^\mu \tilde{p}^\nu$, then the matching leads to
\begin{equation}
-\kappa\,p\wedge \tilde{p} = \kappa \mu^2 \dfrac{1-z^2}{2z} = -\dfrac{2\sqrt{\pi}}{\sqrt{N}} \mt{T}(B_0|B_0\pi_0;z),\label{MatEq}
\end{equation}
in this expression we assume that $p$ and $\tilde{p}$ (momenta of incoming and outgoing $B$ mesons) are on-shell. The sign in the RHS is chosen to make it consistent with our $z\to 1$ limit (\ref{LamConst}). We cannot use this expression in the Lagrangian directly as we need a covariant expression for $z$. Note that making $z$ momentum-dependent would add extra time derivatives to the interaction term and hence disrupt the Hamiltonian, so there is no reason for this procedure to work. Nevertheless, as we will see it works in the heavy quark limit.

In the heavy quark limit when we can assume that the momentum transfer is much smaller than the heavy quark mass $M$, e.g. $1-z \ll 1$:
\begin{equation}
    \tilde{p} \wedge p  = \mu^2\dfrac{1-z^2}{2z} \to M \w,\quad M\to \infty,
\end{equation}
where $M$ is the heavy quark mass and $\w = M(1-z)$ as before. We then have a simple expression $\w = \frac{\tilde{p}\wedge p}{M}$, so after using the large $M$ scaling (\ref{HLTVert}) the matching should imply
\begin{equation}
    \kappa\,p\wedge \tilde{p} = \dfrac{2M \sqrt{\pi}}{\sqrt{N}}\,T_{0|0,0} \lb \dfrac{p\wedge q}{M} \rb,\label{HLEffVert}
\end{equation}
where $q = p-\tilde{p}$ is the pion 2-momentum in covariant formalism. Note that $p\wedge q$ can be negative, while $T_{0|0,0}(\w)$ is defined only for $\w > 0$. However the parity invariance of the effective action implies the natural extension $T_{0|0,0}(-\w) = -T_{0|0,0}(\w)$ as $p\wedge q$ changes sign under the parity transformation.

In order to test (\ref{HLEffVert}) let us compute the correction to the $B$-meson mass using covariant formalism where it is simply given as a one-loop $B$ propagator correction. The above vertex (\ref{HLEffVert}) leads to the following expression:
\begin{equation}
    \delta \mu^2 = \dfrac{i M^2}{N} \ili \dfrac{d^2 q}{\pi} \dfrac{ T_{0|0,0}^2 \lb \frac{p\wedge q}{M} \rb}{(-2pq + q^2 + i\ep)(q^2-\mu_0^2 + i\ep)},\label{SigEff}
\end{equation}
where the external momentum $p$ is on-shell and $q$ is the virtual pion momentum. Due to Lorentz-invariance we can assume $p^0 = M$\footnote{The difference between $M$ and $\mu$ would lead to a contribution vanishing in $M\to \infty$ limit} and $p^1 = 0$, then $pq = M q^0$, $p\wedge q = M q^1$. In the large $M$ limit we can neglect $q^2$ compared to $2pq$ and
\begin{equation}
    \dfrac{1}{-2pq +i\ep} = -i\pi \delta(2 p q) -\mt{P} \dfrac{1}{2pq}.
\end{equation}
The principal value part would vanish as the rest of the integrand is symmetric under $q\to -q$, hence we find:
\begin{equation}
    \delta \mu^2 = -\dfrac{M}{2N} \ili_{-\infty}^\infty dq^1 \dfrac{T_{0|0,0}^2(q^1)}{\lb q^1\rb^2+\mu_0^2} = -\dfrac{M}{N} S_{0|0,0}.
\end{equation}
We see that it exactly coincides with the $B_0\pi_0$ contribution from the full theory analysis (\ref{HLcorr}). The vertex function imposes a natural cutoff on $q$ as we assumed. We can also use the simple cutoff approximation from Subsection \ref{NumResSec} which yields
\begin{equation}
    \kappa_{\text{appr}} = \dfrac{2\sqrt{\pi}}{\sqrt{N}} \theta \lb \mt{S}_0 - \left|\dfrac{p\wedge q}{M}\right| \rb,\label{LamApprox}
\end{equation}
giving an explicit cutoff to $\kappa$. As we found before, this approximation works rather well even when $\mu_0 \neq 0$.

In the finite $M$ case this matching procedure fails: as we mentioned, a momentum-dependent vertex would significantly modify the Hamiltonian. Nevertheless, even the low-energy coupling (\ref{LamConst}) has important connections with other known results, which we will discuss in the next subsection.

\subsection{Exact coupling and WZW model}
Besides the heavy quark limit, there is another interesting case: the double chiral limit $M,m\to 0$. In general, low energy dynamics of $SU(N)$ $\text{QCD}_2$ with $N_f$ flavors of massless fundamental quarks is described by the $U(N_f)$ WZW model at level $N$ \cite{Witten:1983ar,Frishman:1992mr,Delmastro:2021otj}. Its action is as follows:
\begin{equation}
    S[g] = \dfrac{N}{8\pi} \ili d^2x  \Tr \lb\p_\mu g \p^\mu g^{-1} \rb + \dfrac{N}{12\pi}\ili_{\mathcal{B}} d^3y\,\lc^{ijk} \Tr \lb g^{-1}\p_i g g^{-1} \p_j g g^{-1} \p_k g  \rb.
\end{equation}
Here $g$ is a $U(N_f)$ matrix and $\mathcal{B}$ is a three-dimensional volume bounded by the $2d$ spacetime. The coefficients of this action are fixed by current algebra considerations: it should coincide with the algebra of the original fermionic currents. The second term coefficient can also be found from anomaly matching, the derivation for a 4d theory can be found in \cite{Wess:1971yu,Witten:1983tw}.

In our case $N_f = 2$ and $g$ can be represented as $g = e^{i F}$ with
\begin{equation}
    F = \dfrac{2 \sqrt{\pi}}{\sqrt{N}} \begin{pmatrix}
        \pi&\bar{B}\\
        B&\eta
    \end{pmatrix},
\end{equation}
where $\eta$ and $\pi$ are real fields and $B$ is a complex field. They correspond to the mesons described in Section \ref{CorrMes}. In the large $N$ limit we can expand the WZW action around $F = 0$. The integrand of the volume term becomes a total derivative, so the integral localizes to the 2d spacetime. In the leading order we get:
\begin{equation}
    \mt{L} = |\p_\mu B|^2  + \dfrac{1}{2}(\p_\mu \pi)^2 + \dfrac{1}{2}(\p_\mu \eta)^2  - \dfrac{2 i \sqrt{\pi}}{\sqrt{N}} \lc^{\mu\nu} \p_\mu \bar{B} \p_\nu B \pi + \dfrac{2 i \sqrt{\pi}}{\sqrt{N}} \lc^{\mu\nu} \p_\mu \bar{B} \p_\nu B \eta.
\end{equation}
We see that $\pi B \bar{B}$ coupling coincides with the previous result (\ref{LamConst}). As we found, this coupling does not actually change if the second quark becomes massive~--- the WZW turns out to be protected.

The mass-independence of $\kappa$ was actually derived in \cite{Grinstein:1994hy} from axial current conservation. The result is as follows:
\begin{equation}
    \kappa = \dfrac{2}{f_\pi}.
\end{equation}
where $f_\pi$ is the pion decay constant defined by
\begin{equation}
    \bra{0} a^\mu(0) \ket{\pi_0(p)} = f_\pi p^\mu.
\end{equation}
Here $a^\mu = \bar{q}(x) \gamma^\mu \gamma^3 q(x)$ is the massless quark axial current with $\gamma^3 = i\gamma^0\gamma^1$. Note that there is an additional factor of $2$ as our normalization is different from one of \cite{Grinstein:1994hy}. The decay constant be expressed in terms of the pion lightcone wave function from the 't Hooft equation \cite{Callan:1975ps,Umeeda:2021llf}:
\begin{equation}
    f_\pi = \sqrt{\dfrac{N}{\pi}} \ili_0^1 \phi_{\pi_0}(x) dx = \sqrt{\dfrac{N}{\pi}},
\end{equation}
which again leads to the same result (\ref{LamConst}).

Finally, it is interesting to note that the effective theory (\ref{EffLagr}) in the chiral limit $\mu_0=0$ looks like an unparticle theory \cite{Georgi:1990um}~--- a CFT\footnote{In our case it is just a massless scalar, but for multiple massless flavors it would be the corresponding WZW model.} coupled to matter with the coupling vanishing in the infrared. As we discussed, the full theory computations from Section \ref{CorrMes} imply that this coupling should also vanish in the ultraviolet like in (\ref{HLEffVert}) or (\ref{LamApprox}).

\section{Threshold states}\label{ThrSt}
So far we focused our considerations on the chiral limit with one of the flavors becoming massless. We found that the corrections to the meson masses (\ref{GenCorr}) remain finite. However, chiral limit is not the only situation when the integrand of (\ref{GenCorr}) can become singular. As we explained at the beginning of Section \ref{CorrMes}, only a quadratic singularity with coinciding roots of the denominator might lead to a divergence of the integral. It occurs when $\mu_L = \mu_{R_1}+\mu_{R_2}$~--- at the decay threshold. Besides, as we mentioned in Subsection \ref{LightMes} we mentioned that in a 2-flavor model with $m = M$ higher orders of perturbation theory might become divergent when $m\to 0$. In this section we will address these situations, starting with the former.

\subsection{Leading-order divergences}
As we mentioned, we consider the threshold case $\mu_L = \mu_{R_1}+\mu_{R_2}$. Assuming for simplicity that $\delta_{ll'} = 0$, the correction coefficient becomes as follows:
\begin{equation}
    \Sigma_{L|R_1,R_2} = -\ili_0^1 dz \dfrac{\mt{T}^2(L|R_1R_2;z)}{(\mu_L z - \mu_{R_1})^2+i\ep}
\end{equation}
The root is $z_\pm=\frac{\mu_{R_1}}{\mu_L}$ and this integral is nonsingular if $T(L|R_1R_2;z_\pm) = 0$. The parity relation (\ref{SymTPar}) implies that it indeed vanishes when $n_L+n_{R_1}+n_{R_2}$ is even like in the $B_n B_n \pi_0$ vertex case we considered before. This essentially means that the decay $L\to R_1R_2$ should be allowed by parity. However when $n_L+n_{R_1}+n_{R_2}$ is odd the vertex function has a local extremum at $z_{\pm}$ as we observed in Subsection \ref{HLlimit}. Therefore in such situations the perturbation theory breaks down. It is important when $\mu_{R_1}+\mu_{R_2}$ is the continuous spectrum threshold as then a bound state which mixes $R_1R_2$ and $L$ might appear such that the mixing does not vanish in large $N$ limit.

As we are dealing with degenerate perturbation theory, we can limit the mass operator diagonalization problem to a subspace of states spanned by $\ket{L}$ and $\ket{R_1R_2}$. As we will see, the contribution to the mass correction coming from this subspace is parametrically larger than the $\sim 1/N$ result of the non-degenerate calculation. An arbitrary state of the subspace with lightcone momentum $P$ can be represented as follows:
\begin{equation}
    \ket{\zeta_{R_1R_2},C_L;P} = C_L \mt{M}^\dagger_{P,L} \ket{0} + \sqrt{P}\ili_0^1dz\,\zeta_{R_1R_2}(z) \mt{M}^\dagger_{zP,R_1} \mt{M}^\dagger_{(1-z)P,R_2} \ket{0}.
\end{equation}
We added the $\sqrt{P}$ factor to the second term to make the inner product momentum-independent:
\begin{equation}
    \braket{\zeta_{R_1R_2},C_L;P}{\tilde{\zeta}_{R_1R_2},\tilde{C}_L;\tilde{P}} = \delta(\tilde{P}-P) \left[ \ili_0^1 dz\,\tilde{\zeta}_{R_1R_2}(z)\bar{\zeta}_{R_1R_2}(z)+ \tilde{C}_L \bar{C}_L \right].
\end{equation}

Now we can write down the equation for the mass operator eigenstates using the Hamiltonian (\ref{PEff}):
\begin{equation}
    \begin{cases}
        \lb\dfrac{\mu_{R_1}^2}{z} + \dfrac{\mu_{R_2}^2}{1-z}\rb \zeta_{R_1R_2}(z) + \dfrac{C_L}{\sqrt{N}} \dfrac{\mt{T}(L|R_1R_2;z)}{\sqrt{z(1-z)}} = \mtf{m}^2 \zeta_{R_1R_2}(z);\\
        \mu_L^2 C_L+\dfrac{1}{\sqrt{N}} \mathlarger{\ili}_0^1 dz\dfrac{\mt{T}(L|R_1R_2;z)\zeta_{R_1R_2}(z)}{\sqrt{z(1-z)}}  = \mtf{m}^2 C_L,
    \end{cases}\label{SchSys}
\end{equation}
where $\mtf{m}^2$ is the mass of the state squared. Let us denote $\Delta = \mu_L^2-\mtf{m}^2$, a bound state would correspond to $\Delta>0$ as it should be below the threshold. Then expressing $\zeta_{R_1R_2}(z)$ in terms of $\mt{T}(L|R_1R_2;z)$ from the first equation and plugging it into the second one we find:
\begin{equation}
    \Delta = \dfrac{1}{N}\ili_0^1 dz \dfrac{\mt{T}^2(L|R_1R_2;z)}{\Delta z(1-z) + \mu_L^2 (z-z_{\pm})^2}.
\end{equation}
Note that the RHS monotonously drops from infinity to zero with an increase of $\Delta$, so there is only one bound state. At large $N$ we expect $\Delta$ to be small, so we can use the following approximation:
\begin{equation}
    \dfrac{1}{\Delta z(1-z) + \mu_L^2 (z-z_{\pm})^2} = \dfrac{\pi\delta(z-z_{\pm})}{ \mu_L\sqrt{\Delta z_{\pm} (1-z_\pm)}},\quad \Delta \to 0.\label{DelAppr}
\end{equation}
As $\mu_L\sqrt{z_{\pm}(1-z_{\pm})} = \sqrt{\mu_{R_1}\mu_{R_2}}$, we get the following expression for $\Delta$:
\begin{equation}
    \Delta = \dfrac{1}{N^{2/3}}\left[ \dfrac{\pi \mt{T}^2(L|R_1R_2;z_\pm)}{\sqrt{\mu_{R_1}\mu_{R_2}}} \right]^{2/3},\quad N\to\infty.\label{DelEnThr}
\end{equation}

We see that there is indeed a bound state with both $\ket{L}$ and $\ket{R_1R_2}$ components, i.e. it has a tetraquark part. The large $N$ behavior of its mass is rather unusual: we have $\mu_L^2 - \mtf{m}^2 \sim 1/N^{2/3}$ unlike a typical $\sim 1/N$ result. There are other corrections to $\mtf{m}^2$ coming from different choices of intermediate states and from the 4-point vertex acting nontrivially on $\ket{R_1R_2}$, but they scale as $1/N$ and hence are subleading. We can also find the probability of this state appearing as a two-meson state $\nu_{R_1R_2}$:
\begin{equation}
    \nu_{R_1R_2} = \dfrac{\int_0^1 dz|\zeta_{R_1R_2}(z)|^2}{|C_L|^2+\int_0^1 dz|\zeta_{R_1R_2}(z)|^2}. 
\end{equation}
Using the first equation of (\ref{SchSys}), we find:
\begin{equation}
    \ili_0^1 dz |\zeta_{R_1R_2}(z)|^2 = \dfrac{|C_L|^2}{N}\ili_0^1 dz \dfrac{z(1-z)\mt{T}^2(L|R_1R_2;z) }{\left[\Delta z(1-z) + \mu_L^2 (z-z_{\pm})^2\right]^2}.
\end{equation}
Here we can use an approximation similar to (\ref{DelAppr}):
\begin{equation}
    \dfrac{1}{\left[\Delta z(1-z) + \mu_L^2 (z-z_{\pm})^2\right]^2} = \dfrac{\pi\delta(z-z_{\pm})}{2 \mu_L[\Delta z_{\pm} (1-z_\pm)]^{3/2}},\quad \Delta \to 0.
\end{equation}
The explicit expression for $\Delta$ leads to a surprising result:
\begin{equation}
    |C_L|^2 = 2\ili_0^1 dz\,|\zeta_{R_1R_2}(z)|^2;\quad \nu_{R_1R_2} = \dfrac{1}{3},\quad N\to\infty.
\end{equation}
It means a threshold state is always $1/3$ 2-meson and $2/3$ 1-meson independently of the quark masses and the explicit form of the vertex function.

To conclude this chapter, let us apply these results to the heavy-light limit from Section \ref{CorrMes}. For a $B_{n}\to B_l\pi_k$ threshold the formula (\ref{DelEnThr}) becomes as follows with use of the vertex function large $M$ scaling (\ref{HLTVert}) and the on-shell roots formula (\ref{OnShHL}):
\begin{equation}
    \Delta = \dfrac{M}{N^{2/3}}\left[ \dfrac{\pi T_{n|lk}^2(\mu_k)}{\sqrt{\mu_{k}}} \right]^{2/3},\quad N\to\infty.
\end{equation}
The threshold occurs when $\xi_n-\xi_l = 2\mu_k$. We see that this result has a correct large $M$ scaling just like the perturbation theory result (\ref{SCorr}). We will consider the case of  $n=1$, $l=k=0$ threshold depicted on Figure \ref{VertFunPlt} (d). It corresponds to $m=m_t=0.3248$, $\mu_0=1.248$, $\xi_0 = 2.214$ and $\xi_1 = 4.711$. When $m > m_t$, $B_1$ becomes stable ($\mu_{B_1} < \mu_{B_0} + \mu_0$): while there are continuous spectrum states starting from $2\mu_0$, which is much less than $B_1$ mass, the theory preserves flavor charge so for $B_1$ decays at least one meson should contain a heavy quark. We get the following result for the threshold state mass correction:
\begin{equation}
    \Delta = 1.95 \dfrac{M}{N^{2/3}}.
\end{equation}
We see that the numerical coefficient here is comparable to $\xi_1$, so at small $N$ there should be a noticeable difference between the threshold state mass and the 't Hooft model result for $B_1$ mass.

\subsection{Naive divergences in the chiral limit}\label{NaivDiv}
Now we will study the effects of the higher-order divergences in the limit $m=M\to 0$. According to our discussion in Subsection \ref{LightMes}, they appear in corrections to $B_0$ mass when we consider chains of intermediate states of the form $B_0 \to B_0\pi_0 \to B_2 \to B_0\pi_0\to\dots\to B_0\pi_0\to B_0$. As before, we then need to restrict the Hilbert space to the span of $\ket{\pi_2}$, $\ket{\pi_0}$ and $\ket{B_0 \pi_0}$:
\begin{equation}
    \ket{\zeta,C_2,C_0;P} = C_0 \mt{M}_{P,B_0}^\dagger\ket{0} + C_2 \mt{M}_{P,B_2}^\dagger\ket{0} + \sqrt{P} \ili_0^1 dz \zeta(z) \mt{M}^\dagger_{zP,B_0} \mt{M}_{(1-z)P,\pi_0}^\dagger\ket{0}.
\end{equation}
Essentially, this way we can perform the necessary resummation of naive $1/N$ corrections as they become large at $m\to 0$. Similarly to Eq. (\ref{SchSys}) we can obtain a system of equations defining the mass square $\mtf{m}^2$:
\begin{equation}
\begin{cases}
    \mu_0^2 C_0 + \dfrac{\sqrt{2}}{\sqrt{N}} \mathlarger{\ili}_0^1 dz \dfrac{\mt{T}(\pi_0|\pi_0\pi_0;z) \zeta(z)}{\sqrt{z(1-z)}} = \mtf{m}^2 C_0;\\
    \mu_2^2 C_2 + \dfrac{\sqrt{2}}{\sqrt{N}} \mathlarger{\ili}_0^1 dz \dfrac{\mt{T}(\pi_2|\pi_0\pi_0;z)\zeta(z)}{\sqrt{z(1-z)}} = \mtf{m}^2 C_2;\\
    \dfrac{\mu_0^2}{z(1-z)} \zeta(z) + \dfrac{\sqrt{2}C_0}{\sqrt{N}} \dfrac{\mt{T}(\pi_0|\pi_0\pi_0;z)}{\sqrt{z(1-z)}} + \dfrac{\sqrt{2}C_2}{\sqrt{N}}\dfrac{\mt{T}(\pi_2|\pi_0\pi_0;z)}{\sqrt{z(1-z)}} = \mtf{m}^2 \zeta(z).
\end{cases} \label{SmallmSys}
\end{equation}
Note that as $m=M$ we can use $\pi_m$ instead of $B_m$ in vertex functions. We added extra $\sqrt{2}$ factors to account for $\eta_0\pi_0$ intermediate states: they lead to an extra factor of $2$ in the correction (\ref{LeadCorrChir}). Solving this system, we get
\begin{equation}
    \mtf{m}^2 = \mu_0^2 + \dfrac{2}{N} \ili_0^1 dz \dfrac{\mt{T}^2(\pi_0|\pi_0\pi_0;z)}{\mtf{m}^2 z(1-z) - \mu_0^2} + \dfrac{4}{N^2} \dfrac{\left[ \mathlarger{\ili}_0^1 dz\dfrac{\mt{T}(\pi_0|\pi_0\pi_0;z)\mt{T}(\pi_2|\pi_0\pi_0;z)}{\mtf{m}^2 z(1-z) - \mu_0^2} \right]^2}{\mtf{m}^2-\mu_2^2 - \dfrac{2}{N} \mathlarger{\ili}_0^1 dz \dfrac{\mt{T}^2(\pi_2|\pi_0\pi_0;z)}{\mtf{m}^2 z(1-z) - \mu_0^2}}.\label{MassP0P2}
\end{equation}
We are interested in the limits $m\to 0$ and $N\to \infty$. We expect $\mtf{m}^2$ to be close to $\mu_0^2 \sim m$. As we show in Appendix \ref{EqMass},
\begin{equation}
    \mt{T}(\pi_0|\pi_0\pi_0;z) = m \mt{T}_0(z),~m\to 0;\quad \mt{T}_0(z) = - \dfrac{\pi^2 + 3 \log (1-z) \log \dfrac{1-z}{z^2} + 6 \Li_2 \dfrac{z}{z-1} - 6\Li_2 z }{\sqrt{3} \pi}.
\end{equation}
The vertex function $\mt{T}(\pi_2|\pi_0\pi_0;z)$ does not vanish, let us denote its $m\to 0$ limit by $\mt{T}_0(\pi_2|\pi_0\pi_0;z)$. 

The second term in (\ref{MassP0P2}) is $\sim {m}/{N}$, it is the standard second-order correction we studied before in Subsection \ref{LightMes}. The third term is more interesting: naively it is $\sim 1/N^2$ and gives the 4-th order correction. However,
\begin{equation}
\frac{1}{N} \ili_0^1 dz \frac{\mt{T}^2(\pi_2|\pi_0\pi_0;z)}{\mtf{m}^2 z(1-z) - \mu_0^2} \sim \frac{1}{Nm}.
\end{equation}
so when $m \ll 1/N$ the denominator of the third term is dominated by it. Note that this expression can be interpreted as the result of a resummation of higher-order corrections. The third term is $\sim m/N$ when $m \ll 1/N$: we see that $1/N^2$ correction turns into $1/N$ one, but it is still suppressed compared to $\mu_0^2$ and no terms that do not vanish at $N\to \infty$ are generated. It is interesting that this term has a different sign compared to the standard second-order contribution. The share of 2-meson part defined by $\zeta$ is always $\sim 1/N$, so the infinite $N$ spectrum is not modified.

This computation is incomplete as any $\pi_k$ with even $k$ can be an intermediate state in the chain. We can generalize our approach by including them:
\begin{equation}
    \ket{\zeta,C_0,\{C_n\};P} = C_0 \mt{M}_{P,B_0}^\dagger\ket{0} + \sum_{n=1}^\infty C_n \mt{M}_{P,B_n}^\dagger\ket{0} + \sqrt{P} \ili_0^2 dz \zeta(z) \mt{M}^\dagger_{zP,B_0} \mt{M}_{(1-z)P,\pi_0}\ket{0}.
\end{equation}
Then we get a system of equations analogous to (\ref{MassP0P2}). Note that when $m\ll 1/N$ we essentially neglect the terms with $C_2$ in the second equation. A similar property would hold for the general system and can be easily resolved. In order to write the solution in compact form, let us define
\begin{equation}
    V_n = \ili_0^1 \dfrac{\mt{T}_0(z) \mt{T}_0(\pi_n|\pi_0\pi_0;z)}{1-z(1-z)},\quad G_{mn} = \ili_0^1 \dfrac{\mt{T}_0(\pi_m|\pi_0\pi_0;z) \mt{T}_0(\pi_n|\pi_0\pi_0;z)}{1-z(1-z)},\quad n,m=1,\dots,\infty.
\end{equation}
Then we get:
\begin{equation}
    \mtf{m}^2 = 
        \dfrac{2\pi}{\sqrt{3}}m - \dfrac{\sqrt{3}}{\pi }\dfrac{m}{N} \mathlarger{\ili}_0^1 dz \dfrac{\mt{T}_0^2(z)}{1-z(1-z)} + \dfrac{\sqrt{3}}{\pi}\dfrac{m}{N} \mathlarger{\sum\limits}_{k,l=1}^\infty V_k G^{-1}_{kl} V_l,\quad m\ll 1/N\ll 1.
\end{equation}
Note that $G_{mn}$ defines a matrix of scalar products of functions $\mt{T}_0(\pi_n|\pi_0\pi_0;z)$ with measure $\frac{dz}{1-z(1-z)}$. It would be natural to assume that they define a complete basis on $(0,1)$: for instance, numerical data suggests that $\mt{T}_0(\pi_n|\pi_0\pi_0;z)$ oscillates and has $n-1$ zeroes. In means that $ \sum_{l=1}^\infty G_{kl}^{-1} V_l$ are actually the Fourier coefficients of $\mt{T}_0(z)$ in this basis and
\begin{equation}
    \sum\limits_{k,l=1}^\infty V_k G^{-1}_{kl} V_l = \ili_0^1 dz \dfrac{\mt{T}_0^2(z)}{1-z(1-z)}.
\end{equation}
Hence the $\sim m/N$ contribution is exactly cancelled.

There is a simple explanation why it happens. Let us set $C_0 = 1$ in (\ref{SmallmSys}). Then $\mtf{m}^2-\mu_0^2 \sim m$ if $\zeta(z) \sim 1$, $m\to 0$. Let us denote this part of $\zeta$ by $\zeta^{(0)}$ From the last equation it can be inferred that $C_2 \sim m$. In the generalized system we would get $C_n \sim m$, so the second equation in the leading order in $m$ implies
\begin{equation}
\ili_0^1 dz \dfrac{\mt{T}(\pi_n|\pi_0\pi_0;z) \zeta^{(0)}(z)}{\sqrt{z(1-z)}} = 0.
\end{equation}
As $\mt{T}(\pi_n|\pi_0\pi_0;z)$ functions form a complete basis, $\zeta^{(0)}(z) = 0$, hence there is no correction of order $O(m)$. Note that the derivation does not depend on the structure of the 3-point vertex and would still work if we add a 4-point contribution $B_0\pi_0 \to B_0\pi_0$ to the third equation assuming that it is suppressed in $m$. In particular, one can add all 3-meson terms that are subleading in $1/N$, restore $\frac{1}{1-1/N^2}$ factors mentioned in subsection \ref{InfNBas}, and the cancellation would still occur. As the result, if we consider at most 2-meson intermediate states, we get:
\begin{equation}
    \mtf{m}^2 = \begin{cases}
        \dfrac{2\pi}{\sqrt{3}}m - 0.559 \dfrac{m}{N},& 1/N \ll m \ll 1;\\\\
        \dfrac{2\pi}{\sqrt{3}}m + O(m^2),& m \ll 1/N \ll 1.
    \end{cases}
\end{equation}
Here we used the result (\ref{SmallMCorr}) from the end of Subsection \ref{LightMes}. We see that corrections from 2-meson states are suppressed in $m$ even in a multi-flavor model.

Finally, it would be interesting to consider intermediate states with more than $2$ mesons intermediate states, which should give $\sim m$ corrections observed in \cite{Anand:2021qnd}. While we do not have a formal proof, it looks like that higher meson number contributions are suppressed in $1/N$ even after applying degenerate perturbation theory. For instance, a chain of states with 3-meson parts that leads to $m$-independent correction is $B_0 \to B_0\pi_0 \to B_0 \pi_0\pi_0 \to B_1 \pi_0 \to B_0\pi_0\pi_0 \to B_0\pi_0 \to B_0$. It is of order $1/N^3$, so if the small $m$ amplification gives $\sim \frac{1}{Nm}$ term in the denominator\footnote{Note that the amplification should be understood in relative sense: the factor $\sim Nm$ it gives is actually small when $m\ll 1/N$, but $m$ disappears if we divide it by $\mu_0^2$} as it happened in (\ref{MassP0P2}), the resulting correction would be $\sim m/N^2$. The numerical results of \cite{Anand:2021qnd} are not enough to confirm that: the correction coming from at most 6-parton states is $2.09\cdot 10^{-4}$, which is consistent with both $m/N^2 = 1.02 \cdot 10^{-4}$ and $m/N = 3.06\cdot 10^{-4}$.

The main result of this subsection is that while the corrections are still suppressed in $1/N$ in $m\to 0$ limit, their structure changes significantly. In particular, $1/N$ expansion in gauge theories has topological structure \cite{tHooft:1973alw}: a given diagram scales with $N$ as $N^{2-2h-L}$. Here $h$ is the genus of the surface which admits the diagram without self-intersections and $L$ is the number of quark loops which serve as boundaries. The fact the we saw some of the corrections turning from $\sim 1/N^2$ into $1/N$ means that this structure is violated when $m\to 0$.

\section{Discussion}\label{Disc}
In this paper we studied $1/N$ corrections to 't Hooft model mass spectrum in various situations. We have found that these corrections increase significantly in the chiral limit but do not exhibit the divergence claimed in \cite{Barbon:1994au}. In the case of 2-flavor model we found that the corrections to heavy-light meson $B_0$ mass are dominated by one particular intermediate state $B_0\pi_0$ when the light quark becomes massless. We also studied the implications for corrections to the lightest meson $\pi_0$ mass and checked that they are consistent with the numerical results of \cite{Anand:2021qnd}. We showed that the bosonization prediction for $\pi_0$ mass unlike the 't Hooft model result does not match with numerics and our findings. Namely, the bosonization result obtained in \cite{Frishman:1987cx,Frishman:1992mr} implies that the corrections to the 't Hooft model prediction are not suppressed with $N$\footnote{It is important that this is the statement about the corrections and not the values themselves. In particular, the infinite $N$ result for the meson mass $\sim 1$ if $\frac{\bar{g}^2 N}{2\pi} = 1$ and the bosonization finite $N$ result $\sim 1/\sqrt{N}$ (e.g. much smaller). It means that if the bosonization result was correct, the corrections should cancel the infinite $N$ mass, i.e. they should be $\sim 1$}. As we explained in the end of Section \ref{CorrMes}, we do not observe such behavior. Besides, no infinite $N$ mesonic spectrum restructuring occurs in the chiral limit. Then we studied the low energy effective theory and found that $1/N$ corrections give the same 3-meson coupling as WZW model. The coupling stays the same even if one of the quarks becomes massive. For the heavy-light limit we constructed a momentum-dependent coupling which exactly reproduces the $1/N$ correction, this coupling vanishes in both IR and UV. After that we used DLCQ to confirm our results and showed that the improved version of \cite{vandeSande:1996mc} is suitable for the small mass case. We then investigated the $2\to 1$ threshold states. We found that when the decay is allowed by parity, the infinite $N$ spectrum changes and has a bound state that is $1/3$ two-meson and $2/3$ one-meson. The corrections show unusual $\sim 1/N^{2/3}$ scaling. Finally, we showed that higher-order $1/N$ corrections in the massless (strong coupling) limit are amplified: some of $1/N^2$ terms turn into $1/N$. We observed some interesting cancellations of the corrections and found that they do not change the infinite $N$ spectrum structure.

It is rather interesting that we observed a significant increase of corrections if the intermediate state includes $\pi_0$. It happens due to the properties of the expression for corrections (\ref{GenCorr}): unless $\mu_L = \mu_{R_1} + \mu_{R_2}$, the masses in the denominator would suppress the result. If $\mu_{R_2} = 0$ this condition is satisfied for any state $\mu_L$ by $R_1 = L$ and the denominator acquires a second-order zero at $z=1$. The singularity is not present as massless states decouple but it still amplifies the corrections: $\left.\mt{T}'(L|R_1R_2;z)\right|_{z=1} = \mu_L^2$ so the mass in the denominator is cancelled. As the denominator of (\ref{GenCorr}) is purely kinematic and massless states in 2d gauge theories decouple \cite{Kutasov:1994xq}, we can conjecture that $1/N$ corrections are noticeable if the theory contains massless modes. The results of \cite{Dempsey:2022uie} support this: the corrections in adjoint $\text{QCD}_2$ which has no massless mode are of order $10^{-2}$. This is similar to the subleading terms in Table \ref{NumRes}.

It would be interesting to study more complicated theories with massless sector. For instance, in theory with $N_f$ flavors of which $N_f-1$ is light and 1 is heavy we would get an effective theory of heavy-light mesons interacting with $U(N_f-1)$ WZW with central charge $\frac{N(N_f-1)^2}{N+1}$ \cite{Delmastro:2021otj} at level $N$ describing the massless modes. The coupling, as we discussed, should vanish in IR and UV. Hence we would get a more complicated example of unparticle theory \cite{Georgi:2007ek}: instead of $B$ mesons interacting with free boson CFT we had for $N_f = 2$. The $1/N$ corrections would be amplified by a factor of $N_f$, so, as is natural to expect from 't Hooft model, the corrections become large when $N_f \sim N$, i.e. when the central charge $\sim N^2$ at $N\to \infty$. Another interesting theory is $\text{QCD}_2$ with one adjoint and $N_f$ fundamental fermions considered in \cite{Dempsey:2021xpf}. The massless sector is described by a coset WZW model with central charge $\sim 3/4 NN_f$ at $N\to \infty$. So it would be interesting to check if the corrections in this case are amplified even at small $N_f$ due to large central charge.

As of now we do not know why the bosonization result does not seem to predict the correct large $N$ behavior, but it would be nice to see if it can be improved. One possible way would be studying the 4-meson vertex in the small quark mass limit and trying to match it with the interaction predicted by the sine-Gordon Lagrangian. One can also use this vertex to investigate $2\to 2$ threshold states and see if 2-meson bound states appear in the spectrum. The improved DLCQ could be a good way to study the finite $N$ theory numerically in addition to Lightcone Conformal Truncation. Besides, it might be helpful to understand if it is possible to study the small mass limit of $1/N$ corrections to $\pi_0$ mass more systematically than using degenerate perturbation theory in $1/N$. 

Finally, while we considered the purely mesonic sector, baryons become important when the quark mass is small. It would be interesting to see if their masses are consistent with sine-Gordon predictions. It might be possible to generalize the operator expansion (\ref{OpExp}) to the nonzero baryon number case. It would allow one to get the large $N$ limit and $1/N$ corrections of baryonic spectrum systematically. One could also study meson-baryon bound states similarly to \cite{Callan:1985hy,Callan:1987xt} with baryons represented as solitons of the mesonic action \cite{KlebanovWIP}, especially since certain couplings between mesons are known exactly.

\section{Acknowledgements}
We would like to thank Igor Klebanov for proposing this problem and guidance throughout the project. We are also grateful to Z.Sun, R.Dempsey, J.Donahue and U.Trittman for valuable discussions and suggestions. This work was supported in part by the Simons Collaboration on Confinement and QCD Strings through the Simons Foundation Grant 917464.

\begin{appendices}
\section{Relations involving color-singlet operators}\label{aux}
In this Appendix we list some auxiliary relations. We start with the algebra of $M$, $B$ and $D$ operators:
\begin{equation}
\begin{aligned}
    \relax[M_{kk',ii'},M^\dagger_{pp',jj'}] &= \delta_{ij}\delta_{i'j'}\delta(k-k')\delta(p-p') - \dfrac{1}{N}\left[\delta_{i'j'}\delta(k'-p') B_{pk,ij} +\delta_{ij}\delta(k-p) D_{p'k',i'j'}\right];\\
    [M_{kk',ii'},B_{pp',jj'}] &= \delta_{ij}\delta(k-p) M_{p'k',j'i'};\quad [M_{kk',ii'},D_{pp',jj'}] = \delta_{i'j}\delta(k'-p) M_{kp',ij'};\\
    [B_{kk',ii'},B_{pp',jj'}] &= \delta_{i'j}\delta(k'-p) B_{kp',ij'} - \delta_{ij'}\delta(k-p')B_{pk',ji'};\\
    [D_{kk',ii'},D_{pp',jj'}] &= \delta_{i'j}\delta(k'-p) D_{kp',ij'} - \delta_{ij'}\delta(k-p')D_{pk',ji'}.\label{MBDAlg}
\end{aligned}
\end{equation}
Then, the kernels $\mt{K}$ we use in the interaction term (\ref{IntPot}) are as follows:
\begin{equation}
\begin{aligned}
\mt{K}_{MM} &=\left[
\dfrac{\delta_{ij}\delta_{i'j'}}{(k-p)^2}
+\dfrac{1}{N}\dfrac{\delta_{ii'}\delta_{jj'}}{ (k+k')^2}
\right]\delta(k+k'-p-p');\\
\mt{K}_{BB} = \mt{K}_{DD} &= \dfrac{1}{2N}\left[
\dfrac{\delta_{ij'}\delta_{ji'}}{(k-p')^2}
+\dfrac{1}{ N}\,\dfrac{\delta_{ii'}\delta_{jj'}}{ (k-k')^2}
\right] \delta(k-k'+p-p');\\
\mt{K}_{MB} &=\dfrac{2}{\sqrt{N}}\left[
\dfrac{\delta_{ij'}\delta_{ji'}}{(k-p')^2}
+\dfrac{1}{N}\dfrac{\delta_{ii'}\delta_{jj'}}{ (k+k')^2}
\right]\,\delta(k+k'+p-p');\\
\mt{K}_{MD} &= -\dfrac{2}{\sqrt{N}}\left[
\dfrac{\delta_{ij}\delta_{i'j'}}{(k+p)^2}
+\dfrac{1}{ N}\dfrac{\delta_{ii'}\delta_{jj'}}{ (k+k')^2}
\right]\delta(k+k'+p-p');\\
\mt{K}_{BD} &=-\dfrac{1}{ N}\left[
\dfrac{\delta_{ij}\delta_{i'j'}}{(k+p)^2}
+\dfrac{1}{N}\dfrac{\delta_{ii'}\delta_{jj'}}{(k-k')^2}
\right]\,\delta(k-k'+p-p').
\end{aligned}
\end{equation}
Finally, it is instructive to add the interaction term of order $1/N$:
\begin{equation}
    \begin{aligned}
        &~P^-_{(1/N)}=\dfrac{g^2 N}{8\pi N} \sum_{i,i',j,j'}\ili dkdk'dpdp'dqdq' \\
    &~\lb\dfrac{M^{0\dagger}_{qq',ij'}M^{0\dagger}_{pk',ji'}M^0_{kk',ii'}M^0_{pp',jj'}}{(q'-p')^2}\delta(q+q'-k-p') +\dfrac{M^{0\dagger}_{kq',ij'}M^{0\dagger}_{qk',ji'}M^0_{kk',ii'}M^0_{pp',jj'}}{(q-p)^2} \delta(q+q'-p-p')- \right.\\
    &-\dfrac{M^{0\dagger}_{q'p',ij'}M^{0\dagger}_{qk',ji'}M^0_{kk',ii'}M^0_{pp',jj'}}{(q-p)^2}\delta(q+q'-k-p) -\dfrac{M^{0\dagger}_{kq',ij'}M^{0\dagger}_{pq,ji'}M^0_{kk',ii'}M^0_{pp',jj'}}{(q'-p')^2} \delta(q+q'-k'-p')\\
    &+ \left.2 \dfrac{M^{0\dagger}_{qk',ji'}M^{0\dagger}_{pq',j'j}M^0_{kk',ii'}M^0_{pp',j'i}}{(q+q')^2}\delta(q+q'-k-p')\rb.
    \end{aligned}\label{4PVert}
\end{equation}

\section{Numerical methods}\label{NumMet}
Here we will describe the numerical methods we use to compute meson masses, 't Hooft wave functions and heavy-light three-point vertex functions. Besides the general 't Hooft equation (\ref{Th-Eq}) we also have its heavy-light limit (\ref{ThEqHl}) which we need to consider separately. Besides as we will discuss later the general scheme that we use for solving 't Hooft equation is not suitable for the vertex function calculation in the chiral limit. Hence, we also provide a separate approach for such case.

\subsection{Massive 't Hooft equation}\label{MassThEq}
We start with the general 't Hooft equation (\ref{Th-Eq}). For simplicity we denote $m_1 = m_i$ and $m_2=m_{i'}$. One way of solving it commonly used in literature is a decomposition of the wave function $\phi(x)$ in terms of some basis of functions. For instance, it is convenient to introduce a variable $\theta \in (0,\pi)$ with $x = \frac{1-\cos \theta}{2}$ and use 
\begin{equation}
f_n(x) = \sin  (n\theta) = 2 \sqrt{x(1-x)} U_{n-1}(1-2x) \label{BFGen}
\end{equation}
as a basis \cite{Brower:1978wm}, where $U_n(x)$ are Chebyshev polynomials of the second kind and $n\ge 1$ is an integer. Of course there are many other possibilities for basis functions, for example Jacobi polynomials \cite{Chabysheva:2012fe,MO1993159}. As we will see, our choice of basis turns out to be convenient for practical calculations. Let us denote the operator in the RHS of the 't Hooft equation acting on the wave function by $\mt{H}_t$. Then the eigenvalue problem in terms of the basis functions $f_n$ becomes as follows:
\begin{equation}
    \mu^2 \sum_{n}  A_{mn} \phi_n = \sum_l B_{ml} \phi_l,\quad A_{mn} = \braket{f_n}{f_m},~B_{mn}= \bra{f_m} \mt{H}_{t} \ket{f_n},\label{ThMatr1}
\end{equation}
here $\phi(x) = \sum_n \phi_n f_n(x)$. The scalar product is defined in the usual way:
\begin{equation}
    \braket{\chi}{\phi} = \ili_0^1 dx\,\chi(x) \phi(x),
\end{equation}
we assume that the wave functions are real as the spectrum is discrete. Note that $A_{mn}$ is nondiagonal for our choice of $f_n(x)$: these functions are orthogonal with respect to $d\theta$ integration measure. 

The next step is to truncate the summation in (\ref{ThMatr1}) at some $N$ and treat it as a matrix eigenvalue problem for $N\times N$ matrices. However, there is a issue: our basis functions do not satisfy the boundary asymptotic behavior (\ref{PhiAsymp}). It becomes especially problematic when one of the masses is small: if i.e. $m_1 = 0$ we expect $\phi(0) \neq 0$, which is not true for $f_n(x)$. Hence, we correct this by adding two extra functions to the basis:
\begin{equation}
    f_{N+1}(x) = x^{\beta_1}(1-x)^{2-\beta_1},\quad f_{N+2}(x) = x^{2-\beta_2}(1-x)^{\beta_2};\quad \pi\beta_j\cot \pi\beta_j = 1-m_j^2.\label{ExtrBF}
\end{equation}
These functions were used in the original paper \cite{tHooft:1974pnl}. They are convenient to use as the relevant principal value integrals can be expressed in terms of elementary functions \cite{prudnikov1986elem}:
\begin{equation}
    \mt{P} \ili_0^1 dy\dfrac{y^{a}(1-y)^{2-a}}{(x-y)^2}  = \dfrac{\pi}{\sin (\pi a)}\left[(2x-a)(1-x)^{1-a}x^{a-1}\cos(\pi a) + 2x + a - 2 \right].
\end{equation}

The integrals for other basis functions are also rather simple:
\begin{equation}
    \mt{P} \ili_0^1dy \dfrac{\sin( n\tilde{\theta})}{(x-y)^2} = -2\pi n\dfrac{\sin(n\theta)}{\sin\theta};\quad y = \dfrac{1-\cos\tilde{\theta}}{2}.
\end{equation}
One can obtain this result by using residue theorem for $z = e^{i\tilde{\theta}}$ as the integrand for $d\tilde{\theta}$ measure is symmetric under $\tilde{\theta} \to -\tilde{\theta}$ so the integral can be extended to the whole period. The principal value integral in complex coordinates can be represented as follows \cite{tHooft:1974pnl,Brower:1978wm}:
\begin{equation}
    \dfrac{\mt{P}}{(x-y)^2} = \text{Re} \dfrac{1}{(x-y-i\ep)^2}.
\end{equation}
Knowing this, it is straightforward to find the matrices $A$ and $B$. We would need some auxiliary integrals:
\begin{equation}
\begin{aligned}
    &~T_{mn} = \ili_{0}^1 dx\,\sin(n\theta) \sin(m\theta)  = \begin{cases}-\dfrac{2mn}{m^4 + (n^2-1)^2-2m^2(1+n^2)},&(n+m)\mod 2=0;\\
    0,&(n+m)\mod 2 = 1,
    \end{cases}\\
    &~J_{mn} = 2\ili_{0}^1 dx\dfrac{\sin(m\theta)\sin(n\theta)}{1-\cos(\theta)} = \ili_0^\pi d\theta\dfrac{\sin(m\theta)\sin(n\theta)\sin(\theta)}{1-\cos\theta},\\
    &~\mt{I}_n(a_1,a_2) = \ili_0^\pi d\theta\, \sin(n\theta) x^{a_1-1} (1-x)^{a_2-1}  =2n B(a_1,a_2)\,_3F_2(1-n,1+n,a_1;a_1+a_2,3/2|1).
\end{aligned}
\end{equation}
We used \cite{prudnikov1986sp} for the last integral, $\,_3F_2$ is the generalized hypergeometric function and $B(a,b) = \frac{\Gamma(a)\Gamma(b)}{\Gamma(a+b)}$ is the beta function. Note that its the integration measure is $d\theta$, using $dx$ would lead to an expression of a similar form as $dx~=~\sqrt{x(1-x)}d\theta$. The integral for $J_{mn}$ can be found using $\theta \to \pi-\theta$ transformation, which gives
\begin{equation}
    J_{mn} = (-1)^{n+m} \ili_0^\pi d\theta\dfrac{\sin(m\theta)\sin(n\theta)\sin(\theta)}{1+\cos\theta}
\end{equation}
When $(n+m) \mod 2 = 0$ adding it to the previous expression for $J_{mn}$ leads to an integral that was already computed in \cite{Brower:1978wm,Lebed:2000gm}:
\begin{equation}
    J_{mn} = \ili_0^\pi d\theta \dfrac{\sin(m\theta) \sin(n\theta)}{\sin\theta} = 2\sum_{l=\frac{|m-n|}{2}+1}^{\frac{m+n}{2}} \dfrac{1}{2l-1} = J_{m-1,n-1} + \dfrac{2}{m+n-1}.
\end{equation}
On the other had, when $(n+m)\mod 2 = 1$ adding the two expression yields:
\begin{equation}
    J_{mn} = \ili_{0}^\pi d\theta \dfrac{\sin(m\theta)\sin(n\theta)\cos(\theta)}{\sin(\theta)} = \dfrac{J_{m,n-1} + J_{m,n+1}}{2}.
\end{equation}
So in the end we get:
\begin{equation}
    J_{mn} = \begin{cases} 2\sum_{l=\frac{|m-n|}{2}+1}^{\frac{m+n}{2}} \dfrac{1}{2l-1},&(m+n)\mod 2 =0\\\\
    \dfrac{J_{m,n+1}+J_{m,n-1}}{2},&(m+n) \mod 2 = 1.
    \end{cases}
\end{equation}

Now we can write down explicit expressions for $A$ and $B$. Let us assume that $n,m = 1,\dots,N$ and treat $N+1$ and $N+2$ indices separately. We also only need to specify the lower triangular parts as the matrices are symmetric. First, for $A$ get:
\begin{equation}
\begin{gathered}
    A_{mn} = T_{mn},\quad A_{N+1,n} = \mt{I}_n(\beta_1+3/2,7/2-\beta_1),\quad A_{N+2,n} = \mt{I}_n(7/2-\beta_2,\beta_2+3/2),\\
    A_{N+1,N+1} = B(2\beta_1+1,5-2\beta_1),\quad A_{N+2,N+1} = B(3+\beta_1-\beta_2,3+\beta_2-\beta_1),\\A_{N+2,N+2} = B(5-2\beta_2,2\beta_2+1).
\end{gathered}
\end{equation}
Next,
\begin{equation}
\begin{aligned}
    &~B_{mn} = \dfrac{n\pi^2}{2} \delta_{mn} +(\alpha_1+              (-1)^{n+m}\alpha_2)J_{mn},\\ 
    &~B_{N+1,n} = \alpha_1\mt{I}_n(\beta_1+1/2,7/2-\beta_1) +\alpha_2 \mt{I}_n(\beta_1+3/2,5/2-\beta_1)+ 2\pi n \mt{I}_n(\beta_1+1,3-\beta_1),\\
    &~B_{N+2,n} = \alpha_1 \mt{I}_n(5/2-\beta_2,\beta_2+3/2) + \alpha_2 \mt{I}_n(7/2-\beta_2,\beta_2+1/2) +2\pi n \mt{I}_n(3-\beta_2,\beta_2+1),\\
    &~B_{N+1,N+1} = (\alpha_2-\alpha_1)B(2\beta_1+1,4-2\beta_1)+2\frac{\alpha_1}{\beta_1}\Big[B(2\beta_1+1,4-2\beta_1) - \sec(\pi\beta_1)B(\beta_1+1,4-\beta_1)\Big]\\
    &+\alpha_1 \sec(\pi\beta_1) B(\beta_1+1,3-\beta_1),\\
    &~B_{N+2,N+1} = (\alpha_2-\alpha_1)B(3+\beta_1-\beta_2,2-\beta_1+\beta_2)\\
    &+2\frac{\alpha_1}{\beta_1}\Big[B(3+\beta_1-\beta_2,2-\beta_1+\beta_2) - \sec(\pi\beta_1)B(3-\beta_2,\beta_2+2)\Big] + \alpha_1 \sec(\pi\beta_1)B(3-\beta_2,\beta_2+1),\\
    &~B_{N+2,N+2} = (\alpha_1-\alpha_2)B(4-2\beta_2,2\beta_2+1)+2\frac{\alpha_2}{\beta_2}\Big[B(4-2\beta_2,2\beta_2+1) - \sec(\pi\beta_2)B(4-\beta_2,\beta_2+1)\Big] \\
    &+\alpha_2 \sec(\pi \beta_2) B(3-\beta_2,\beta_2+1).
\end{aligned}
\end{equation}
Here $\alpha_j = m_j^2 - 1$. Note that some of the matrix elements have ${1}/{\beta_j}$ factors, but the limit $\beta_j \to 0$ is still well-defined. Now the approximate meson masses and wave functions can be found from the truncated form of equation (\ref{ThMatr1}). This method converges very fast with $N$, however the convergence becomes worse for very small but nonzero $m_j$ and for very large $m_j$. Besides, numerical calculations show that the extra functions (\ref{ExtrBF}) are not necessary for good precision if $m_{1,2}>0.5$~--- e.g. when the endpoint behavior of them becomes more smooth compared to the other functions $f_n(x)$. We use $N=20$ for light mesons with $1>m_j \ge 0.01$ and $N = 100$ for heavy mesons with $100\ge m_j \ge 10$, this ensures at least 4 digits of precision for meson masses.

It is important that we used the $x$ space scalar product so our matrices are symmetric. Without the extra basis functions (\ref{ExtrBF}) it is more convenient to use $d\theta$ scalar product \cite{Brower:1978wm,Lebed:2000gm} as it makes $A$ diagonal, although $B$ is then not symmetric. However it leads to singular matrix elements with the extra functions if one of the masses is zero, so that approach is not applicable as we are interested in the chiral limit.

Another issue is that this approach fails to work when $m_1 = m_2 = 0$. It can be fixed by projecting out the zero mode $\phi(x) = 1$. Nevertheless, the endpoint behaviour of numerical wave functions would still be
\begin{equation}
    \phi(x) \sim \begin{cases}
        \sqrt{x},&x\to 0\\
        \sqrt{1-x},&x\to 1
    \end{cases}
\end{equation}
as it follows from (\ref{BFGen}). This leads to $\mt{T}(L|L\pi_0;z) \sim \sqrt{z}$, $z\to 0$ for the vertex function (\ref{TVertGen}) involving massless $\pi_0$. As the result the correction integral (\ref{SigSing}) being divergent\footnote{As this divergence is a numerical artefact, by increasing $N$ one can move the integration region contributing to the divergence closer to $z = 1$.}, so our method is not suitable for the chiral limit. Note that it was used in \cite{Barbon:1994au}, so this is likely a source of the divergence in that work. For $\pi_0$ we can simply use the exact wave function $\phi(x)=1$ and in our case $L = B_n$ is a heavy-light meson. If its wave function also has the $\sqrt{1-x}$ behavior at $x\to 1$ the correction integral would still diverge. It means that for the heavy quark limit we need an approach that does not have this boundary behavior problem at $x\to 1$. We will describe it in the next subsection.

\subsection{Heavy-light 't Hooft equation}\label{HLThEq}
As in the previous subsection, we decompose the solution $g(t)$ of the equation (\ref{ThEqHl}) using some basis of functions $h_n(t)$: $g(t) = \sum_n g_n h_n(t)$. A convenient orthonormal choice is
\begin{equation}
    h_n(t) =  2^{\beta+1/2}\sqrt{\dfrac{\Gamma(n)}{\Gamma(n+2\beta)}} t^{\beta}  e^{-t} L^{(2\beta)}_{n-1}(2t),\quad \pi\beta\cot\pi\beta = 1-m^2,
\end{equation}
where $L^a_n(x)$ are generalized Laguerre polynomials and $n \ge 1$ is an integer. It gives the correct $t\to 0$ behavior and simultaneously is smooth at $t=0$ when $m\to 0$. It does not automatically satisfy the $g'(0) = 0$ at $m=0$ property of the original equation, which would lead to $T_{n|lk}(\w) \sim \w\log \w$ for a numerical approximation of the heavy-light vertex function (\ref{HLTVert}). Nevertheless, the correction integral (\ref{SSing}) would still be convergent, so this choice of basis should be good enough.

As before, we need to find the matrix elements to obtain the matrix equation. We denote the operator in the RHS of (\ref{ThEqHl}) by $\mt{H}_{HL}$, then
\begin{equation}
    \xi g_l = \sum_n B_{ln}^{(HL)} g_n,\quad B_{ln}^{(HL)} = \bra{h_l} \mt{H}_{HL} \ket{h_n}
\end{equation}
as the basis functions are orthonormal. We again use the standard scalar product:
\begin{equation}
    \braket{g}{\tilde{g}} = \ili_0^{\infty} dt\, g(t) \tilde{g}(t).
\end{equation}
It is hard to evaluate the matrix elements directly, but we can represent $h_n(t)$ as a sum of more elementary functions:
\begin{equation}
    h_l(t) = \sum_{n} L_{ln} u_{n}(t),\quad u_n(t) = \dfrac{t^{n+\beta-1}}{\Gamma(n+\beta)}.\label{AuxFuncU}
\end{equation}
The coefficients $L_{ln}$ follow from the direct formula for $L_{m-1}^{2\beta}(2t)$:
\begin{equation}
    L_{ln} = \begin{cases}
    (-1)^{n-1}2^{n+\beta-1/2}\dfrac{\sqrt{\Gamma(l+2\beta)\Gamma(l)}\Gamma(n+\beta)}{\Gamma(l-n+1)\Gamma(n+2\beta)\Gamma(n)},&n\leq l;\\
    0,&n>l.
        \end{cases}
\end{equation}

The matrix elements for $u_l$ are more straightforward to compute. The most problematic part is the principal value part. A convenient way to deal with it is to use a Fourier transform as the kernel in the principal value integral corresponds to $-\pi |k|$ in momentum space \cite{Freese:2022ipu}. The Fourier image of $u_n(t)$ is as follows:
\begin{equation}
    \hat{u}_n(k) = \ili_0^{\infty} dt\, u_n(t) e^{-i k t} = \dfrac{1}{(1+ik)^{n+\beta}}.
\end{equation}
Then we find:
\begin{equation}
\begin{aligned}
    &~\mt{P} \ili_0^\infty dt ds\, \dfrac{u_l(t) u_n(s)}{(t-s)^2} = -\dfrac{1}{2}\ili_{-\infty}^{\infty}dk \dfrac{|k|}{(1+ik)^{l+\beta}(1-ik)^{n+\beta}}\\
    &=\dfrac{_2F_1(2,n+\beta;l+n+2\beta|2) +\,_2F_1(2,l+\beta;l+n+2\beta|2)}{2(l+n+2\beta-2)(l+n+2\beta-1)},
\end{aligned}
\end{equation}
where we used an integral from \cite{prudnikov1986elem}. The matrix elements for $u_n(t)$ are then as follows:
\begin{equation}
\begin{aligned}
    ~&\tilde{B}^{(HL)}_{ln} = \bra{u_l}\mt{H}_{HL}\ket{u_n} = \dfrac{(l+n+2\beta-2)(l+n+2\beta-1)-4}{2^{l+n+2\beta}(l+n+2\beta-2)(l+n+2\beta-1)B(l+\beta,n+\beta)}\\
    &-\dfrac{_2F_1(2,n+\beta;l+n+2\beta|2) +\,_2F_1(2,l+\beta;l+n+2\beta|2)}{2(l+n+2\beta-2)(l+n+2\beta-1)} + \dfrac{4m^2 \Gamma(l+n+2\beta-2)}{2^{l+n+2\beta}\Gamma(l+\beta)\Gamma(n+\beta)}.
\end{aligned}
\end{equation}
The first two terms are singular when $l=n=1$ and $m=0$ but their difference is finite:
\begin{equation}
    \left.\tilde{B}_{11}^{(HL)}\right|_{m=0} = \dfrac{1}{4} + \log 2.
\end{equation}
We can now express $B^{(HL)}$:
\begin{equation}
    B^{(HL)} = L \tilde{B}^{(HL)} L^T.
\end{equation}

Numerical calculations show fast convergence of $\xi$ with increasing truncation size $N$. We use $N=10$ which is enough for 5 digits of precision for $\xi_0$ when $m>0.01$ or when $m=0$. Similarly to the previous subsection, there is a drop in precision for very small but nonzero $m$.

\subsection{Massless 't Hooft equation}\label{MasslThEq}
As we noted before, in the chiral limit we can use the exact wave function for $\pi_0$ to avoid divergence of the correction integral $S_{n|n0}$. The other integrals $S_{n|lk}$, $k>0$ would not be divergent even if $\phi^{(k)}(x) \sim \sqrt{x}$, $x\to 0$ due to the denominator of (\ref{SCorr}). It means in principle we can use the method of Subsection \ref{MassThEq} for the excited pion states in the chiral limit. Nevertheless, if $m_1=m_2 = 0$ there is a simpler way to choose the basis functions so we well describe it in this subsection.

We will use an orthonormal system of Legendre polynomials:
\begin{equation}
    e_n(x) = \sqrt{2n+1}P_n(2x-1),
\end{equation}
$n$ is an integer. We will use $n\ge 1$ as $e_0(x) = 1$ corresponds to the massless wave function $\phi^{(0)}(x)$ which is orthogonal to the other wave functions. This way we will only get the excited states. The matrix elements $B^{(C)}_{mn}$ for these functions were computed in \cite{Anand:2020gnn}. Assuming $n \ge m$, we get:
\begin{equation}
\begin{aligned}
    &~B^{(C)}_{mn} = \bra{e_m} \mt{H} \ket{e_n}\\ 
    &=2\sqrt{(2m+1)(2n+1)}\left[ \psi\lb \frac{n+2}{2} \rb + \psi \lb \frac{n+1}{2} \rb - \psi \lb \frac{n-m+1}{2} \rb - \psi \lb \frac{n+m+2}{2}\rb   \right],\\
\end{aligned}
\end{equation}
where $\psi(x) = \frac{\Gamma'(x)}{\Gamma(x)}$ is the digamma function.

The meson masses can then be found from the eigenvalue equation as usual:
\begin{equation}
    \mu^2 \phi_l = \sum_n B^{(C)}_{ln} \phi_n.
\end{equation}
We use size $N=20$ as the truncation size, it gives 5 digits of precision for the first 5 excited states. Besides, the coefficients $\phi_n$ vanish for either odd or even $n$ depending on the parity of the state according to our discussion in Subsection \ref{VertProp} (parity is opposite to the wave function behavior under $x\to 1-x$).

\subsection{Vertex function}\label{HLVFunc}
Finally, let us describe the method of numerical computation of the heavy-light vertex function $T_{n|lk}(\w)$ and the usual vertex function $\mt{T}(L|R_1R_2;z)$. We start with the former. First, the formula (\ref{HLTVert}) is not very suitable for numerics as it contains a double integral. In order to rewrite it in a more suitable form, let us define some auxiliary functions:
\begin{equation}
    \Phi^{(k)}(x) = \ili_0^1 dy \dfrac{\phi^{(k)}(y)}{(x+y)^2},\quad G^{(l)}(t) = \ili_0^\infty ds \dfrac{g^{(l)}(s)}{(t+s)^2}.
\end{equation}
They can be easily computed numerically as we can explicitly compute the integrals of the relevant basis functions from the previous subsections. We start with $f_n(x)$:
\begin{equation}
\begin{aligned}
    \ili_0^{1} dx \dfrac{x^\beta (1-x)^{2-\beta
}}{(x+y)^2} &= \pi \lb \dfrac{y}{1+y} \rb^{\beta-1} (\beta+2y)\csc(\pi \beta) + \pi(\beta-2y-2)\csc(\pi\beta);\\
    \ili_0^{1} dx \dfrac{\sin(n\theta)}{(x+y)^2} &= (-1)^{n+1} \dfrac{2\pi n e^{-n \tilde{\theta}}}{\sinh\tilde{\theta}},\quad x = \dfrac{1-\cos\theta}{2},~y = \dfrac{\cosh\tilde{\theta}-1}{2}.
\end{aligned}
\end{equation}
Next, using \cite{prudnikov1986elem} we get:
\begin{equation}
    \ili_0^{\infty} ds \dfrac{u_n(s)}{(t+s)^2} = t^{n+\beta-2}\Psi(n+\beta;n+\beta-1|t),
\end{equation}
where $\Psi(a;b|x)$ is the confluent hypergeometric function \cite{GradRyz}. We can then find the similar integral for $h_n(t)$ using (\ref{AuxFuncU}). For the chiral limit basis we have \cite{prudnikov1986sp}:
\begin{equation}
    \ili_0^1 dx \dfrac{e_n(x)}{(x+y)^2} = (-1)^n\dfrac{\sqrt{2n+1}(n+1)B(n+1,n+1)}{t^{n+2}}~_2F_1\lb n+2,n+1;2n+2\left|-\frac{1}{y}\right.\rb.
\end{equation}

Now we can use these functions to rewrite $T_{n|lk}(\w)$:
\begin{equation}
\begin{aligned}
    &~T_{n|lk}(\w) = I^{(1)}_{n|lk}(\w) - I^{(2)}_{n|lk}(\w);\\
    &~I^{(1)}_{n|l,k}(\w) = \ili_0^{\infty} \dfrac{dx}{x}\lb x g^{(n)}[\w(1+x)] g^{(l)}(\w x) \Phi^{(k)}(x) - \dfrac{g^{(n)}(\w) g^{(l)}(0) \phi^{(k)}(0)}{x+1}\rb,\\
    &~I^{(2)}_{n|l,k}(\w) = \ili_0^{1} \dfrac{dx}{x}\Big( \w x g^{(n)}[\w(1-x)] G^{(l)}(\w x) \phi^{(k)}(x) - {g^{(n)}(\w) g^{(l)}(0) \phi^{(k)}(0)}\Big).
\end{aligned}
\end{equation}
We obtained the second term in each integrand by adding and subtracting $g^{(n)}(\w) g^{(l)}(0) \phi^k(0)$ from the numerator of (\ref{HLTVert}). We do not need them when $m\neq 0$ and they indeed vanish, but for $m = 0$ they ensure that these integrals converge. Otherwise only their difference would be finite in the chiral limit. This way we can compute $T_{n|lk}(\w)$ as a difference of two single integrals.

Finally, the general vertex function $\mt{T}(L|R_1R_2;z)$ can be expressed in the same way:
\begin{equation}
\begin{aligned}
    \mt{T}(L|R_1R_2;z) &= \mt{I}_1(L|R_1R_2;z)-\mt{I}_2(L|R_1R_2;z);\\
    \mt{I}_1(L|R_1R_2;z) &=  \ili_0^{\frac{1}{1-z}} \dfrac{d x}{x} \lb x z\phi_{L_1}(z-xz(1-z))\phi_{R_1}(1 - x(1-z)) \Phi_{R_2}(zx) - \dfrac{\phi_{L_1}(z) \phi_{R_1}(1) \phi_{R_2}(0)}{z x+1}\rb;\\
    \mt{I}_2(L|R_1R_2;z) &= \ili_0^{\frac{1}{z}} \dfrac{dx}{x} \lb x (1-z) \phi_{L_1}(z+z(1-z)x) \phi_{R_2}(z x) \Phi_{R_1}(-1-(1-z) x)  - \dfrac{\phi_{L_1}(z) \phi_{R_1}(1) \phi_{R_2}(0)}{x(1-z)+1} \rb.\\
\end{aligned}
\end{equation}
It can be evaluated using above integrals.

\section{Properties of the vertex function near chiral limit}
In this section we study various asymptotic properties the vertex function of $\mt{T}(L|L\pi_0;z)$.

\subsection{Asymptotic behavior at endpoints}\label{DerCalc}
We start with the endpoint derivative in the chiral limit $m_{r_2}=m_{r_2'}=m_{l'}=0$. From (\ref{TAsymp}) it follows that in this case as $\phi_{\pi_0}(y) = 1$ we get:
\begin{equation}
\begin{aligned}
&~\mt{T}'(L|L\pi_0;1) = -\ili_0^1 dx\dfrac{\phi_L^2(1-x)-\phi_L(1) \phi_L(1-x)}{x^2} \\
&=-\ili_0^1 dx\dfrac{[\phi_L(1-x)-\phi_L(1)]^2}{x^2} - \phi(1)\ili_0^1 dx \dfrac{\phi_L(1-x)-\phi_L(1)}{x^2}.
\end{aligned}
\end{equation}
Let us denote the first term here by $J_1$ and the second by $J_2$. They are both convergent as in Subsection \ref{VertProp} we showed that $\phi_L'(1) = 0$. The expression (\ref{1Limit}) yields
\begin{equation}
    J_2 = (\mu_L^2-m_l^2)\phi_L^2(1).
\end{equation}

In order to compute $J_1$ we can use the following trick. Let us multiply the 't Hooft equation (\ref{Th-Eq}) for $\phi_L$ by $x\phi_L'(x)$ and integrate by $dx$ from $0$ to $1$. Assuming that $\phi_L(x)$ has unit norm after integration by parts we find:
\begin{equation}
    \dfrac{\mu_L^2\left[\phi_L^2(1)-1\right]}{2} = \dfrac{m_l^2\left[\phi_L^2(1)-\phi_L^2(0)\right]}{2} + \mt{P} \ili_0^1 dx dy \dfrac{[\phi_L(x)-\phi_L(y)] x \phi_L'(x)}{(x-y)^2}.
\end{equation}
Here we kept $\phi_L(0)$ for generality but $m_l \phi_L(0) = 0$ for any $m_l$ due to the corresponding asymptotic behavior. The principal value integral can be simplified by integrating by parts with respect to $dy$:
\begin{equation}
\begin{aligned}
    &~\mt{P} \ili_0^1 dx dy \dfrac{[\phi_L(x)-\phi_L(y)] x \phi_L'(x)}{(x-y)^2} = -\dfrac{1}{2}\ili_0^1 dx \lb \dfrac{x}{1-x} \dfrac{d}{dx}[\phi(x)-\phi(1)]^2 + \dfrac{d}{dx} [\phi(x)-\phi(0)]^2 \rb \\
    &+\mt{P} \ili_0^1 dx dy  \dfrac{x \phi'(x) \phi'(y)}{x-y} = -\dfrac{J_1}{2}.
\end{aligned}
\end{equation}
In the second line we integrated by parts in the single integrals and used the variable exchange symmetry of the principal value integral to replace $x\to \frac{x-y}{2}$ so
\begin{equation}
    \mt{P} \ili_0^1 dx dy  \dfrac{x \phi'(x) \phi'(y)}{x-y} = \dfrac{1}{2} \ili_0^1 dx dy\,\phi'(x) \phi'(y) = \dfrac{1}{2}[\phi(1)-\phi(0)]^2.
\end{equation}
Therefore we find:
\begin{equation}
    J_1 = \mu_L^2 - (\mu_L^2-m_l^2)\phi_L^2(1).
\end{equation}
Adding up $J_1$ and $J_2$ leads to
\begin{equation}
    \mt{T}'(L|L\pi_0;1) = J_1+J_2 = \mu_L^2.
\end{equation}

Now let us turn on the light quark mass $m_{r_2} = m_{r_2'} = m \ll 1$ and study how it changes the $z\to 1$ behavior. It is convenient to still denote the chiral limit wave functions as $\phi(x)$, using $\tilde{\phi}(x)$ for nonzero $m$. Then we can define a variation $\delta \phi(x) = \tilde{\phi}(x) - \phi(x)$. While computing the variation of the vertex function (\ref{TVertGen}) $\delta \mt{T}(L|R_1R_2;z)$ we would get 3 terms $\delta^{(i)} \mt{T}(L|R_1R_2;z)$, $i=1,2,3$, containing $\delta \phi_{R_1}$, $\delta \phi_{R_2}$ and $\delta \phi_L$ respectively. In our situation $L$ and $R_1$ are the same state, but it is convenient to distinguish them here. For the first two terms we have $\delta^{(1,2)}\mt{T}(L|R_1R_2;z) \sim 1-z$, $z\to 1$ as $\phi_L$ stays the same in these terms and we only needed  the properties of $\phi_L$ to prove such behavior in Subsection \ref{VertProp}. It is reasonable to assume that the coefficients at $1-z$ would be of order $m$.

The situation with $\delta^{(3)}\mt{T}(L|R_1R_2;z)$ is different. The vertex function that we get only by changing $\phi_L$ to $\tilde{\phi}_L$ is as follows:
\begin{equation}
    \tilde{\mt{T}}^{(3)}(L|L\pi_0;z) = \ili_0^{z/(1-z)} d\tau \ili_0^1 dy \dfrac{\phi_{L}(y) \{ \tilde{\phi}_L[z-\tau(1-z)]  - \tilde{\phi}_L[ z+y(1-z)] \}}{\lb y + \tau \rb^2},\label{T3Var}
\end{equation}
where we defined $\tau = \frac{z}{1-z}x$. We are interested in $z\to 1$ behavior of this expression, from the wave function asymptotics (\ref{PhiAsymp}) we obtain:
\begin{equation}
    \tilde{\phi}_L(x) = C_L (1-x)^\beta,~x\to 1;\quad \beta = \dfrac{\sqrt{3}}{\pi}m + O(m^2).
\end{equation}
As it should be consistent with $m\to 0$ limit which turns $\tilde{\phi}_L$ into $\phi_L$ we expect $C_L = \phi_L(0) + O(m)$. Using this expression for $\tilde{\phi}_l$ in (\ref{T3Var}) when $z\to 1$ leads to
\begin{equation}
    \tilde{\mt{T}}^{3}(L|L\pi_0;z) = \beta \phi_L^2(0)(1-z)^\beta \ili_0^\infty d\tau \ili_0^1 dy \dfrac{\log(1+\tau)-\log(1-y)}{(y+\tau)^2} + O(m^2),\quad z\to 1.
\end{equation}
The double integral can be computed exactly, its value is $\pi^2/3$. We see that the contribution from $\delta^{(3)} \mt{T}(L|L\pi_0;z)$ is dominating at $z\to 1$ compared to other variations as they do not have $(1-z)^\beta$ terms.

If we assume $|m \log(1-z)| \ll 1$ and $z \ll 1$, the above relations imply
\begin{equation}
    \tilde{\mt{T}}(L|L\pi_0;z) = -\mu_L^2(1-z) + \dfrac{\pi}{\sqrt{3}} \phi_L^2(0)\,m.
\end{equation}
For the heavy-light vertex after using the large $M$ scaling from Subsection \ref{HLlimit} this translates to
\begin{equation}
    T_{n|n0}(\w) = -\w + \dfrac{\pi}{\sqrt{3}} g^{(n)2}(0)\,m.
\end{equation}

\subsection{Equal mass case behavior}\label{EqMass}
Now let us assume that all quark masses are equal to $m \to 0$ and $L$ is a ground state, so we are actually computing $\mt{T}(\pi_0|\pi_0\pi_0;z)$. Then the integrand of (\ref{TVertGen}) is nonzero only near the endpoints for $\phi_L$: otherwise it is almost constant. Therefore we are only interested in near endpoint behavior. Then we can expand $\phi_L(x) = 1 + \beta \log x + \beta \log(1-x) + O(\beta^2)$ and find in the leading order
\begin{equation}
    \mt{T}(\pi_0|\pi_0\pi_0;z) = \beta \dfrac{1-z}{z}\ili_0^1 dx dy \dfrac{\log \left[ \dfrac{(1-x)z [1-(1-x)z]}{[z+y(1-z)][1-z-y(1-x)]} \right]}{\lb x + y \dfrac{1-z}{z} \rb^2},\quad m\to 0.
\end{equation}
This integral can be computed analytically and using the expression for $\beta$ from the previous subsection we find:
\begin{equation}
    \mt{T}(\pi_0|\pi_0\pi_0;z) = - \dfrac{\pi^2 + 3 \log (1-z) \log \dfrac{1-z}{z^2} + 6 \Li_2 \dfrac{z}{z-1} - 6\Li_2 z }{\sqrt{3} \pi} m,\quad m\to 0.
\end{equation}
Here $\Li_2 z$ is the polylogarithm \cite{prudnikov1986integrals}.
\end{appendices}
\printbibliography
\end{document}